\documentclass[12pt,a4paper]{article}

\usepackage{epsfig}
\usepackage{amsmath,amsfonts,amssymb}
\usepackage{cite}
\usepackage{colortbl}
\usepackage{pifont}

\providecommand{\openone}{\leavevmode\hbox{\small1\kern-3.8pt\normalsize1}}

\newcommand{\ntrk}{N_\text{trk}}

\begin{document}

\begin{center}
\begin{Large}
{\bf Stealth multiboson signals}
\end{Large}

\vspace{0.5cm}
J.~A.~Aguilar--Saavedra \\
\begin{small}
{ Departamento de F\'{\i}sica Te\'orica y del Cosmos, 
Universidad de Granada, \\ E-18071 Granada, Spain} \\ 

\end{small}
\end{center}

\begin{abstract}
We introduce the `stealth bosons' $S$, light boosted particles with a decay $S \to AA \to q \bar q q \bar q$ into two daughter bosons $A$, which subsequently decay into four quarks that are reconstructed as a single fat jet. Variables that measure the two-pronged structure of fat jets, which are used for diboson resonance searches in hadronic or semi-leptonic final states, classify the jets produced in stealth boson decays
as QCD-like --- actually, for these variables they may seem more background-like than the QCD background itself. The number of tracks in those jets can also be, on average, much higher than for the fat jets arising from the hadronic decay of boosted $W$ and $Z$ bosons. Therefore, these elusive particles are hard to spot in standard searches. Heavy resonances decaying into two such stealth bosons, or one plus a $W/Z$ boson, could offer an explanation for the recurrent small excesses found in hadronic diboson resonance searches near an invariant mass of 2 TeV.
\end{abstract}


\section{Introduction}

Small excesses around an invariant mass of 2 TeV appear in no less than five different searches for diboson resonances decaying hadronically, performed by the ATLAS and CMS Collaborations at the Large Hadron Collider (LHC), and with energies of 8 and 13 TeV. Although none of these excesses is statistically significant on its own, it is difficult to regard them as a mere coincidence. Yet, their interpretation as a background shaping or a new physics signal is difficult too. 

The largest excess, of $3.4\sigma$, was found by the ATLAS Collaboration in a search using the full Run 1 datset at 8 TeV~\cite{Aad:2015owa}, and sparked a great interest. The excess appeared in the (non-independent) samples tagged as $WW$, $WZ$ and $ZZ$, but it was largest for the $WZ$ event selection. Before, a mild excess had been found by the CMS Collaboration~\cite{Khachatryan:2014hpa}, though the maximum significance was below $2\sigma$, and located at slightly smaller invariant masses. The first ATLAS analysis of Run 2 data with 3.2 fb$^{-1}$ at 13 TeV~\cite{ATLAS:2015msj}, did not show any hint of an excess with the nominal $WZ$ event selection but relaxing the boson tagging criteria for fat jets a third bump appeared, again around 2 TeV, in one of the control distributions (the dijet invariant mass distribution without an upper cut on the number of tracks per jet). The significance of this bump was later estimated to be of $2.4\sigma$~\cite{Aguilar-Saavedra:2016xuc}. When the analysis was updated with a luminosity of 15.5 fb$^{-1}$~\cite{ATLAS:2016yqq}, a bump at 2 TeV appeared also with the  nominal $WZ$ selection, with a significance around $2\sigma$. The fifth 2 TeV bump has appeared recently in the CMS search with the full 2016 dataset, using a luminosity of 35.9 fb$^{-1}$~\cite{Sirunyan:2017acf}. Its significance is around $2\sigma$. No excess at this mass was observed in earlier analyses with 2.6 fb$^{-1}$~\cite{CMS:2015nmz} and 12.9 fb$^{-1}$~\cite{CMS:2016mwi}. A very recent analysis~\cite{Aaboud:2017eta} by the ATLAS Collaboration with 36.7 fb$^{-1}$ shows no excess at this invariant mass.

\begin{figure}[t]
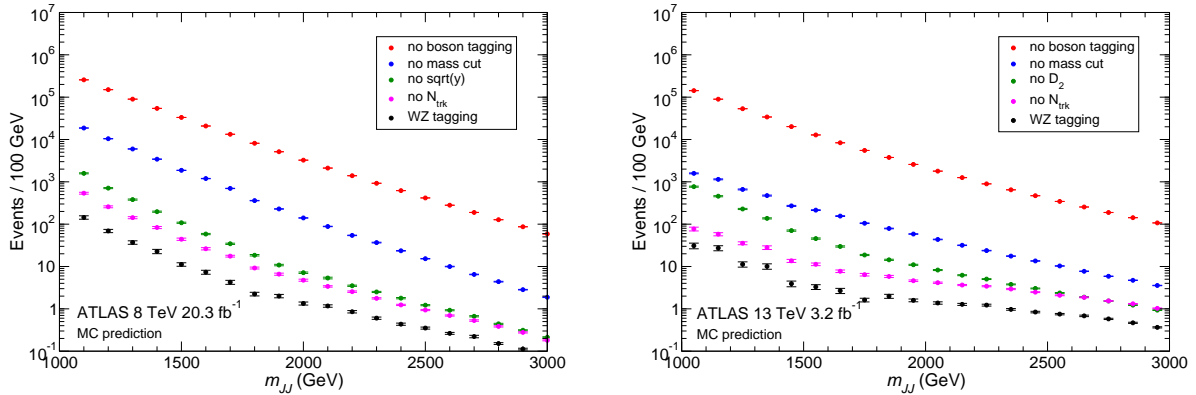

\begin{center}
\begin{tabular}{ccc}
\includegraphics[height=5.2cm,clip=]{Figs/QCD_A-8TeV} & \quad &
\includegraphics[height=5.2cm,clip=]{Figs/QCD_A-13TeV}
\end{tabular}
\caption{Monte Carlo predictions for the QCD dijet background, with the event selections for the ATLAS Run 1 (left) and Run 2 (right) hadronic diboson searches, from Ref.~\cite{Aguilar-Saavedra:2016xuc}. The black lines  correspond to the full $WZ$ boson tagging criteria described in Refs.~\cite{Aad:2015owa,ATLAS:2015msj}: jet mass cuts, $\sqrt y$ (at Run 1) or $D_2$ (at Run 2) substructure tagging, and number of tracks per jet. For the blue, green and pink lines two out of these three tagging criteria are applied. The red lines correspond to the prediction without any boson tagging. The error bars in the points represent the Monte Carlo uncertainty. }
\label{fig:QCD_A}
\end{center}
\end{figure}

Before addressing other explanations, it is worthwhile discussing in detail the possibility that the 2 TeV bumps are caused by some shaping of the dijet invariant mass distribution of the SM background. It is known (see for example Ref.~\cite{Dasgupta:2013ihk}) that the various methods of jet grooming, such as mass-drop filtering~\cite{Butterworth:2008iy}, trimming~\cite{Krohn:2009th}, pruning~\cite{Ellis:2009me} and soft drop~\cite{Larkoski:2014wba} alter the shape of the mass distribution of jets resulting from quarks and gluons. For example, the jet trimming algorithm used by the ATLAS Collaboration in Run 2 searches~\cite{ATLAS:2016yqq} gives rise to a kink in the distribution at masses $m_J \simeq 80$ GeV for jets of radius $R=1.0$ and transverse momenta $p_T = 1$ TeV. In this situation, it is conceivable that a jet mass selection around the $W/Z$ masses could induce some feature in the background distribution around $p_T = 1$ TeV, which could be reflected at dijet invariant masses $m_{JJ} \sim 2$ TeV. On the other hand, the ATLAS Run 1 search uses mass-drop filtering, for which this kink is not present, and the CMS Collaboration uses jet pruning and soft drop in their Run 1 and Run 2 analyses, respectively. 
It is then questionable that the same background shaping would appear at 2 TeV in all these analyses, using not only different jet grooming methods but also different jet substructure tagging. An explicit calculation of the QCD dijet distribution with the ATLAS Run 1 and Run 2 jet tagging criteria~\cite{Aguilar-Saavedra:2016xuc} showed that a kink around $m_{JJ} = 1.7$ TeV appears when the jet mass and tagging criteria are applied (see also Ref.~\cite{Martin:2016jdw}), as it can be seen in Fig.~\ref{fig:QCD_A}. But it shows no hint of a bump around 2 TeV. As it will be shown in the following, the jet tagging criteria used by the CMS Collaboration in Run 2 searches do not produce a 2 TeV bump either. Consequently, in this work we will discard the possibility of a significant background shaping, although further studies on this issue are welcome.

Regarding new physics interpretations, it was early pointed out~\cite{Aguilar-Saavedra:2015rna} that the $3.4\sigma$ ATLAS excess at 8 TeV was unlikely to result from a $VV$ diboson resonance, $V=W,Z$, as the searches in the semi-leptonic final states~\cite{Aad:2015ufa,Khachatryan:2014gha} did not exhibit any deviation from the standard model (SM) prediction. The proposal in Ref.~\cite{Aguilar-Saavedra:2015rna} was of a $VVX$ triboson resonance, that is, a resonance $R$ undergoing a cascade decay $R \to V Y \to VVX$ yielding two SM bosons plus an additional particle $X$, with $Y$ being an intermediate resonance. The presence of an extra particle $X$ would dramatically decrease the efficiency of such a signal in the searches in semileptonic modes, as it was later confirmed in Ref.~\cite{Aguilar-Saavedra:2016xuc} with a more detailed analysis. The reason is that the mentioned Run 1 searches in semileptonic modes, as well as the more recent ones with Run 2 data~\cite{ATLAS:2016npe,ATLAS:2016cwq,CMS:2017mrw,CMS:2017sbi,CMS:2016pfl}, are highly optimised for the kinematics of diboson resonances produced back-to-back in the transverse plane. Moreover, the event selection criteria often veto the presence of extra particles near the decay products of the boson with leptonic decay, which obviously dampens the efficiency for such a signal. As an alternative explanation for the lack of a positive signal in the semileptonic decay channels, it was proposed~\cite{Allanach:2015blv} that the two hadronically decaying `bosons' identified as two fat jets are not actually massive SM gauge bosons but smuons of similar mass, decaying into two quarks via $R$-parity violating interactions, and the heavy 2 TeV resonance is a sneutrino. That is, the observed `bosons' simply do not have leptonic decay modes.

In Ref.~\cite{Aguilar-Saavedra:2017iso} it has been shown that a wide bump on a smoothly-falling distribution that cannot be predicted from simulations is quite difficult to detect in narrow resonance searches: the bump can easily be absorbed in the background normalisation. This is indeed the case for diboson resonance searches in hadronic final states, where the background is obtained from a fit to data in the signal region, assuming some smooth functional form. This feature may partially explain why excesses are not seen in the searches that have a smaller dataset. Still, current new physics interpretations are undermined by an apparent inconsistency among the size of the excesses. For example, in Ref.~\cite{Aguilar-Saavedra:2016xuc} a variety of triboson signals that accommodated the ATLAS Run 1 excess was examined in order to give predictions for other analyses, in particular for the hadronic Run 2 CMS search with 2.6 fb$^{-1}$. The predictions were compatible with the null experimental result but in the latest dataset~\cite{Sirunyan:2017acf}, with a luminosity 13 times higher, a larger excess should have appeared. The same reckoning is expected to apply to a simple signal such as the one proposed in Ref.~\cite{Allanach:2015blv}, with two particles from a resonance decay, each decaying into a quark pair. A possible solution to this puzzle is suggested by a comparison of the different hadronic searches, focusing not on the number of observed events, but rather on the {\it expected\ } QCD background at dijet invariant masses around 2 TeV, and its dependence on the boson tagging criteria~\cite{charla}:
\begin{itemize}
\item CMS analyses use a subjettiness ratio $\tau_{21}$~\cite{Thaler:2010tr} to quantify the likeliness that the jets have a two-pronged structure. For the so-defined `high purity' (HP) jets, the criterion has strenghthened from $\tau_{21} \leq 0.5$ in the Run 1 analysis~\cite{Khachatryan:2014hpa}, to $\tau_{21} \leq 0.45$~\cite{CMS:2015nmz}, $\tau_{21} \leq 0.4$~\cite{CMS:2016mwi}, and $\tau_{21} \leq 0.35$~\cite{Sirunyan:2017acf}, as the dataset has increased. To give an example, the expected background near 2 TeV for the high purity dijet sample is similar in the latest analysis, with 35.9 fb$^{-1}$, and the previous one with 12.9 fb$^{-1}$. This clearly shows that the requirements on jets are much tighter. 
\item For the ATLAS analyses the comparison between Run 1 and Run 2 is not easy, because in Run 1 a cut $\sqrt y \geq 0.45$ on the $y$ variable~\cite{Butterworth:2008iy} measuring the subjet momentum balance is used, whereas in Run 2 it is replaced by a momentum-dependent cut on the so-called $D_2$ function~\cite{Larkoski:2014gra}. Nevertheless, by a naive parton luminosity scaling of the expected background around 2 TeV, it is seen that the boson tagging criteria for the Run 2 nominal selection~\cite{ATLAS:2015msj} are around one order of magnitude stronger.\footnote{Scaling the expected background of two events per 100 GeV in the bin around 2 TeV in the Run 1 analysis by a factor of 15 to account for the increased $gg$ parton luminosity, and multiplying by $3.2~\text{fb}^{-1}/\;20.3~\text{fb}^{-1}$, one would expect a background of 7 events in the Run 2 analysis with 3.2 fb$^{-1}$, whereas the expected background is around 0.7 events.} The latest ATLAS search~\cite{Aaboud:2017eta} implements a new jet mass definition and a reoptimisation of the jet mass windows and $D_2$ cut, whose details are not publicly available. We will not cover that analysis here.
\end{itemize}

Getting together the above arguments, namely (i) the persistence of the 2 TeV bumps, (ii) the unlikely possibility of a background shaping, (iii) the apparent inconsistency of the size of the bumps, an obvious question arises: {\it May it be possible to have new physics signals giving fat jets whose substructure is background-like?} If this were the case, those signals would be more and more suppressed as the jet tagging requirements are tightened, and we might have an explanation for that apparent inconsistency. To answer this question, which arises in the context of the 2 TeV anomaly but whose consequences go beyond the interpretation of potential LHC excesses, is the motivation of the present work. And, as we show in this paper, the answer is affirmative. The cascade decay of a very boosted particle $S$ with a mass $M_S \sim M_V$ into two lighter particles $A$ that subsequently decay each into two quarks,
\begin{equation}
S \to A \, A  \to q \bar q q \bar q \,,
\label{ec:stealth1}
\end{equation}
gives a single fat jet with a mass consistent with the $W$ and $Z$ masses. (The particles $S$ and $A$ can  be for instance a new scalar $H_1^0$ and pseudo-scalar $A^0$, respectively, in models with an extended scalar sector.)
And, depending on the $A$ mass $M_A$, the fat jet originated in the decay of $S$ may seem more background-like than the QCD background itself, when one considers jet subjettiness measures such as $\tau_{21}$ or $D_2$. Therefore, it is appropriate to denote these $S$ bosons as `stealth bosons'. An alternative is to have a heavier $S$ and with one of its decay products being a vector boson $V = W,Z$. One theoretically well-motivated possibility is
\begin{equation}
S \to Z \, A  \to q \bar q q \bar q \,,
\label{ec:stealth2}
\end{equation}
with $M_S \gtrsim M_Z + M_A$ and the four quarks merging into a single jet. (Again, $S$ and $A$ can be the new scalar $H_1^0$ and pseudo-scalar $A^0$ in models with an extended Higgs sector.) Often the jet grooming algorithm will completely eliminate the decay products of $A$, so that the groomed jet mass may be close to the $Z$ mass even if $S$ is heavier.

Stealth bosons with a mass $M_S \sim M_V$ and decaying as in Eq.~(\ref{ec:stealth1}) are the main focus of our analysis. The results obtained apply, at least qualitatively, to the decay chain of Eq.~(\ref{ec:stealth2}) as well. In section~\ref{sec:2} we consider the diboson-like decay of a heavy resonance $R$ into two such stealth bosons, $R \to SS$, and compare the resulting variables with those for a true diboson resonance $Z' \to W W$. There, our statement that those signals can be more background-like than the QCD background will be apparent, with the explicit example of CMS and ATLAS diboson resonance searches. In section~\ref{sec:3} we study a diboson-like decay $R \to V S$ of a resonance into a SM boson and a stealth boson, and in section~\ref{sec:4} a triboson-like cascade decay $R \to V Y \to V V S$, with $Y$ an intermediate resonance. As an example of these cascade decays, as well as for the Monte Carlo simulations, we use the multiboson signals that arise in left-right models~\cite{Aguilar-Saavedra:2015iew}, but our results will not be limited to such specific examples. We discuss our results in section~\ref{sec:5}. Although the novel `stealth boson' signatures introduced in this work are motivated by the 2 TeV excesses, we point out that they are interesting on their own. Such signatures are relatively hard to spot over the QCD background with the standard diboson resonance searches, and therefore they are so far quite unexplored. Related details are discussed in three appendices. The possible shaping of the QCD background with the CMS event selection criteria is investigated in appendix~\ref{sec:a0}. A side effect of the multi-pronged structure of stealth bosons is the fact that standard grooming algorithms often fail to recover the true stealth boson mass from the resulting fat jet. This issue is discussed in appendix~\ref{sec:a}, with a comparison of the results of different grooming algorithms.
Appendix~\ref{sec:b} is devoted to a brief exploration of the decay chain in Eq.~(\ref{ec:stealth2}).


\section{Diboson-like $R \to SS$ decays}
\label{sec:2}

\begin{figure}[htb]
\begin{center}
\includegraphics[height=3cm,clip=]{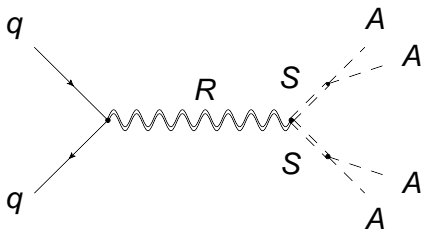}
\caption{Sample diagram for a $R \to SS$ signal.}
\label{fig:diag1}
\end{center}
\end{figure}

The $R \to SS$ decay depicted in Fig.~\ref{fig:diag1} actually yields a quadriboson final state, but will resemble a diboson resonance if $S$ is much lighter than $R$. Such a signal can arise in left-right models if CP is violated in the scalar sector, so that the CP-odd and CP-even states mix, or in simpler $Z'$ extensions of the SM with additional scalars.
We take in our simulations $R = Z'$ with $M_R = 2$ TeV, $S = H_1^0$ with $M_S = 100$ GeV and consider $A = A^0 \to b \bar b$ decays in `high mass' and `low mass' benchmarks, with $M_A = 40$ and 20 GeV, respectively. (As these two scalars do not couple to $WW$ or $ZZ$, they can be quite light and yet evade current searches; moreover, their couplings to quarks can be small, so that they are mainly produced from the decay of heavier particles.)
The relevant Lagrangian is implemented in {\scshape Feynrules}~\cite{Alloul:2013bka} in order to generate events with {\scshape MadGraph5}~\cite{Alwall:2014hca} using the universal Feynrules output~\cite{Degrande:2011ua}. Event generation is followed by hadronisation and parton showering with {\scshape Pythia~8}~\cite{Sjostrand:2007gs}. The detector response is simulated with {\scshape Delphes 3.4}~\cite{deFavereau:2013fsa}, and for the jet reconstruction and analysis {\scshape FastJet 3.2}~\cite{Cacciari:2011ma} is used. The QCD dijet background is generated by slicing the phase space in dijet invariant mass intervals of 100 GeV between 700 and 3.5 TeV, generating $1.5 \times 10^5$ events in each $m_{JJ}$ interval and recombining these samples with a weight proportional to their cross section. 

The CMS hadronic searches use jets with a large radius $R=0.8$ reconstructed with the anti-$k_T$ algorithm~\cite{Cacciari:2008gp}, referred to as AK8 jets, as vector boson candidates. A soft drop grooming algorithm~\cite{Larkoski:2014wba} is performed on the AK8 jets to eliminate contamination from initial state radiation, multiple interactions and pile-up. The most recent analysis selects events with two jets $J_1$, $J_2$ (ordered by transverse momentum) having pseudorapidities $|\eta| \leq 2.5$, pseudorapidity difference $|\Delta \eta| \leq 1.3$, transverse momentum $p_T \geq 200$ GeV and invariant mass $m_{JJ} \geq 1.05$ TeV. These kinematical criteria, which are usually referred to as `topological selection', are quite similar for previous CMS analyses. Jets are considered as $W$-tagged if they satisfy a condition on $\tau_{21}$ specified below and their mass is in the range $65-85$ GeV, and $Z$-tagged if their mass is in the range $85-105$ GeV. 

\begin{figure}[t]
\begin{center}
\includegraphics[height=5.25cm,clip=]{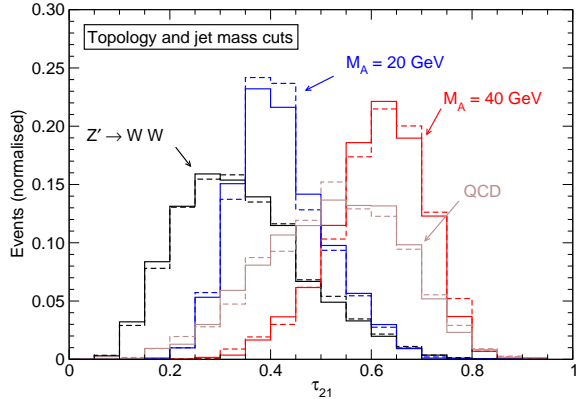}
\caption{$N$-subjettiness ratio $\tau_{21}$ for $R \to SS$ and $Z' \to W W$ signals and the QCD dijet background. Solid (dashed) lines correspond to the leading (subleading) jet.}
\label{fig:XX-CMS1}
\end{center}
\end{figure}

\begin{figure}[t]
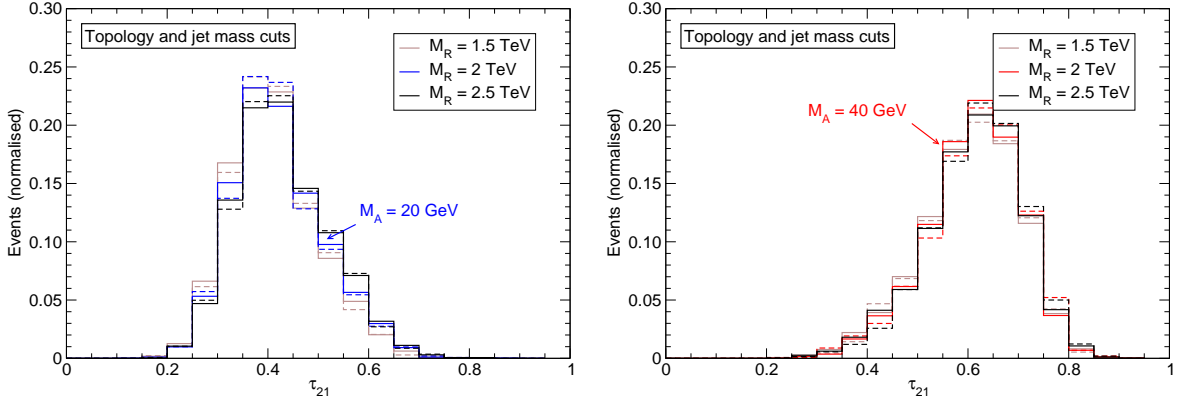

\begin{center}
\begin{tabular}{cc}
\includegraphics[height=5.25cm,clip=]{Figs/tau21-M-A20.eps} &
\includegraphics[height=5.25cm,clip=]{Figs/tau21-M-A40.eps}
\end{tabular}
\caption{$N$-subjettiness ratio $\tau_{21}$ for $R \to SS$ signals with different heavy resonance masses. Solid (dashed) lines correspond to the leading (subleading) jet.}
\label{fig:tau21-M}
\end{center}
\end{figure}

The $N$-subjettiness ratio variable $\tau_{21} = \tau_2^{(1)} / \tau_1^{(1)}$ of the plain (ungroomed) jets is used to enrich the sample with diboson candidates. A low-purity (LP) category is defined for jets with $\tau_{21} \leq 0.75$, and a HP category with $\tau_{21} \leq 0.45 , 0.4, 0.35$, depending on the particular analysis considered. Events are classified as HP if they have two HP jets, and LP if they have one HP and one LP jet. 
The $\tau_{21}$ distributions for the leading and subleading jets for a $Z' \to W W$ sample with $M_{Z'} = 2$ TeV and the QCD background are shown in Fig.~\ref{fig:XX-CMS1}, for jets within the mass window $65-105$ GeV and fulfiling the above acceptance criteria. (For the background, dijet invariant masses in the interval $[1.7-2.3]$ TeV after simulation are considered, in order to compare with the signal on a similar kinematical range.) The distributions have a shape quite similar to the one obtained by the CMS Collaboration~\cite{Sirunyan:2017acf}, also with a maximum in the $0.25-0.3$ bin for the $Z' \to WW$ signal and in the  $0.5-0.55$ bin for the QCD background. This, in particular, shows that the $\tau_{21}$ variable is adequately modeled by the fast simulation. In contrast, the distributions for $R \to SS$ lean towards higher values of $\tau_{21}$ and are background-like, especially for $M_A = 40$ GeV. These distributions are quite independent of the heavy resonance mass, which determines how boosted the stealth bosons are. We have generated $R \to SS$ signals for additional masses $M_R = 1.5,2.5$ TeV and show the results in Fig.~\ref{fig:tau21-M}, for $M_A = 20$ GeV (left) and $M_A = 40$ GeV (right). The normalised distributions for $M_R = 1.5$, $2$, $2.5$ TeV are almost identical. On the other hand, the distributions do depend on the stealth boson mass itself, though this dependence is less interesting for the present study because diboson resonance searches fix jet mass windows around $M_W$ or $M_Z$.

The background-like behaviour of stealth bosons is reflected in a significant efficiency drop as the $\tau_{21}$ cut is tightened. For example, for the HP $WZ$ selection (two HP jets, one $W$-tagged and the other $Z$-tagged)
the efficiencies for the $WW$ diboson, QCD background and $SS$ stealth diboson signals are reduced as follows,\footnote{For the benchmark $Z' \to W W$ diboson signal the efficiency is larger with the $WW$ selection. Still, we give our results for the $WZ$ selection for better comparison with previous work, and also because we have assumed that the new boson $S$ is heavier than the $Z$ boson. In any case, the overall efficiency for $WW$, $WZ$ or $ZZ$ mass tagging for the signals does not affect our arguments.}
\begin{align*}
& W W: && 0.053 \to 0.024 && ( \times \; 0.46) \,, \\
& \text{QCD:} && 4.6 \times 10^{-5} \to 5.9 \times 10^{-6} && ( \times \; 0.13) \,, \\
& M_A = 20~\text{GeV}: && 0.039 \to 0.0031 && ( \times \; 0.08) \,,
\end{align*}
when strengthening the cut from $\tau_{21} \leq 0.45$ to $\tau_{21} \leq 0.35$. (For $M_A = 40$ GeV the efficiency is essentially zero already for $\tau_{21} \leq 0.45$.) This is precisely the `anomalous' behaviour anticipated for this type of signals in the introduction. Note that for the QCD background the absolute efficiency values are not meaningful because they depend on the parton-level sample. In particular the sample generated here has dijet masses $m_{jj} \geq 700~\text{GeV}$, while a cut $m_{JJ} \geq 1.05~\text{TeV}$ is applied in the topological selection. Instead, the relevant issue here is that while the signal-to-background ratios $S/B$ and $S/\sqrt{B}$ increase for a potential $WW$ signal when strengthening the cut, they do not for a stealth boson signal.

\begin{figure}[t]
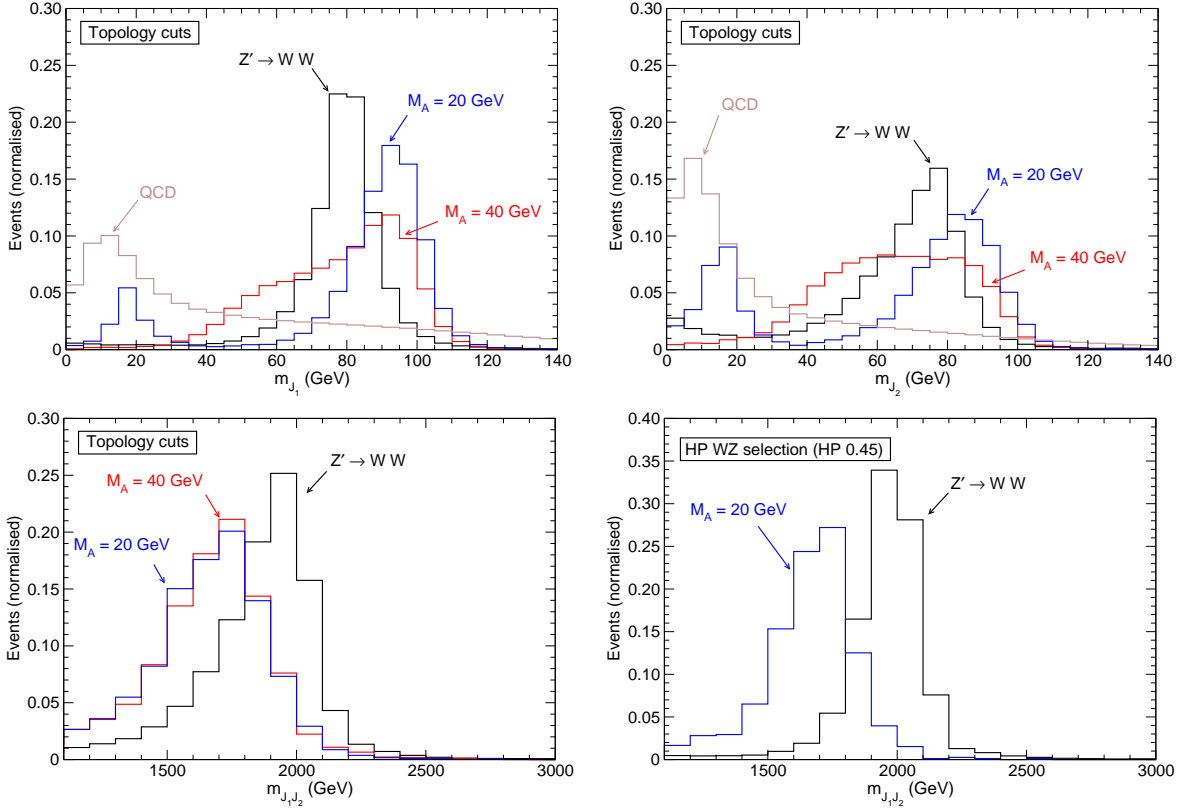

\begin{center}
\begin{tabular}{cc}
\includegraphics[height=5.25cm,clip=]{Figs/mJ1_c-Zp.eps} &
\includegraphics[height=5.25cm,clip=]{Figs/mJ2_c-Zp.eps} \\
\includegraphics[height=5.25cm,clip=]{Figs/mJJ_c-Zp.eps}  &
\includegraphics[height=5.25cm,clip=]{Figs/mJJ_ct-Zp.eps} 
\end{tabular}
\caption{Jet and dijet mass distributions for $R \to SS$ and $Z' \to W W$ signals and the QCD dijet background for the CMS analysis.}
\label{fig:XX-CMS2}
\end{center}
\end{figure}

The jet mass distributions are also of interest. They are shown, after topological selection but before the boson tagging, in the top panels of Fig.~\ref{fig:XX-CMS2}, for the leading and subleading jets. While for the lower mass benchmark the boson masses are adequately reconstructed after the grooming, this is not the case for the higher mass, though a sizeable fraction of events still fall within the $W$ or $Z$ tagging mass windows. (A comparison of the jet masses with different grooming algorithms is presented in appendix~\ref{sec:a}.)
The dijet invariant mass distribution before boson tagging is wider than for a true diboson resonance, and peaks at lower invariant masses, as shown in the bottom left panel of Fig.~\ref{fig:XX-CMS2}. After boson tagging (right panel), the distribution for $M_A = 20$ GeV is still wide while for $M_A = 40$ GeV the simulated signal does not pass the boson tagging. For this plot, we consider HP jets as those with $\tau_{21} \leq 0.45$. We point out that, as shown in Ref. \cite{Aguilar-Saavedra:2017iso}, for a wide bump the location of the maximum deviation with respect to the background-only hypothesis is not located at the maximum of the `signal' distribution, but may be shifted $100-200$ GeV to higher dijet masses. Thus, these distributions would still give an apparent excess near 2 TeV.

We now study the features of the $R \to SS$ signals under the ATLAS Run 2 analyses. The ATLAS Collaboration uses wider jets with radius $R = 1.0$, reconstructed with the anti-$k_T$ algorithm, and trimmed~\cite{Krohn:2009th} to eliminate contamination. As topological selection, events must contain two fat jets, the leading one with $p_T \geq 450$ GeV and the subleading one with $p_T \geq 200$ GeV, both within $|\eta| \leq 2.0$ and with a small rapidity separation $|\Delta y_{12}| \leq 1.2$. The dijet invariant mass $m_{JJ}$ must be larger than 1 TeV. At variance with the CMS analyses, a 
 transverse momentum balance cut is here introduced, $(p_{T1}-p_{T2})/(p_{T1}+p_{T2}) < 0.15$.

\begin{figure}[t]
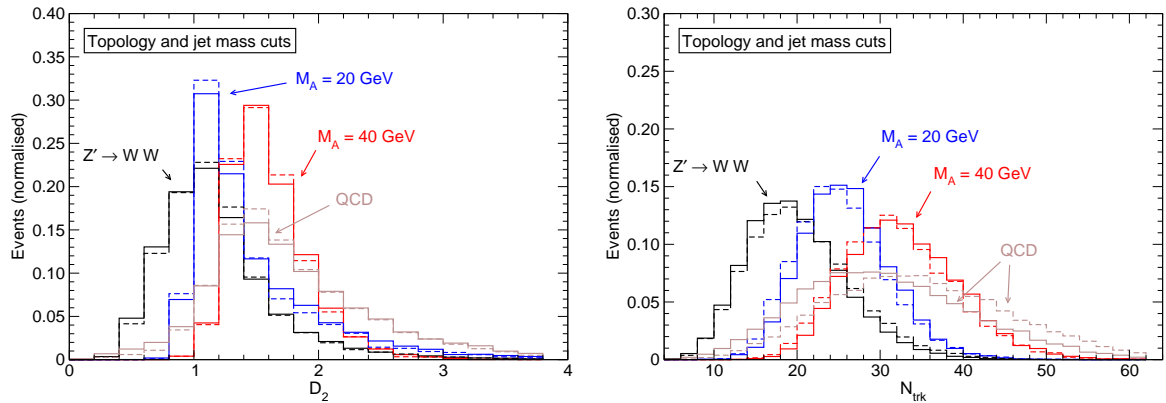

\begin{center}
\begin{tabular}{cc}
\includegraphics[height=5.25cm,clip=]{Figs/D2-Zp.eps} &
\includegraphics[height=5.25cm,clip=]{Figs/Ntrk-Zp.eps}
\end{tabular}
\caption{$D_2$ variable (left) and number of tracks (right) for $R \to SS$ and $Z' \to W W$ signals and the QCD dijet background. Solid (dashed) lines correspond to the leading (subleading) jet.}
\label{fig:XX-ATLAS1}
\end{center}
\end{figure}

The $D_2^{(\beta=1)}$ variable, abbreviated throughout this paper as $D_2$, is used to characterise the two-pronged substructure, and a $p_T$-dependent upper cut is imposed~\cite{ATLASD2}, approximately $D_2 \leq 1 + 0.8 \, (p_T - 300) /1200$ with $p_T$ in GeV. An upper cut $\ntrk < 30$ is also placed on the number of tracks with $p_T \geq 0.5$ GeV in the plain jets, pointing to the primary vertex. Jets satisfying both criteria  are tagged as $W$ or $Z$ candidates if their mass is within an interval of $\pm 15$ GeV around the expected resonance peak. Notice that a jet can simultaneously  be tagged as $W$ and $Z$, therefore the $WW$, $WZ$ and $ZZ$ samples are not disjoint.

We present the $D_2$ distribution for the different signals and the background after topology and mass cuts in Fig.~\ref{fig:XX-ATLAS1} (left). For $Z' \to W W$ and the background the distributions are similar to the ones obtained by the ATLAS Collaboration~\cite{dibosonXtra}, slightly shifted to smaller $D_2$ values. For $R \to SS$ the distributions are background-like, especially for $M_A = 40$ GeV. This is expected from what we have observed for the analogous $\tau_{21}$ variable used by the CMS Collaboration. The application of the $D_2$ requirement on both jets has a 50\% efficiency for the $Z' \to W W$ diboson signal, 24\% for $R \to SS$ in the low-mass benchmark and only 6\% in the high-mass benchmark. 

The distribution of the number of tracks $\ntrk$ for the plain jets is shown in the right panel of Fig.~\ref{fig:XX-ATLAS1}. For a true diboson signal $Z' \to W W$, our results are close to the ones obtained by the ATLAS Collaboration, with a slightly smaller number of tracks predicted by our simulation. For $R \to SS$ with $M_A = 20$ GeV, the distribution has a maximum at the $25-26$ bin, close to the maximum of the QCD background. For $M_A = 40$ GeV the signals have, on average, more tracks than the background itself. For the $WZ$ selection, the requirement $\ntrk < 30$ reduces the efficiency of the signals and background as
\begin{align*}
& W W: && 0.18 \to 0.16 && ( \times \; 0.89) \,, \\
& \text{QCD:} && 6.2 \times 10^{-5} \to 2.0 \times 10^{-5} && ( \times \; 0.32) \,, \\
& M_A = 20~\text{GeV}: && 0.10 \to 0.065 && ( \times \; 0.65) \,, \\
& M_A = 40~\text{GeV}: && 0.017 \to 0.0024  && ( \times \; 0.14) \,.
\end{align*}
Again, the absolute value of the background efficiency is not relevant because it depends on the parton-level cuts in the sample; rather, the relative efficiency with and without the $\ntrk < 30$ requirement is the quantity of interest.
We see that a requirement of a small number of tracks leads to an additional suppression of this type of signals, besides the one from the $D_2$ tagging, which may even decrease the signal-to-background ratio.

\begin{figure}[t]
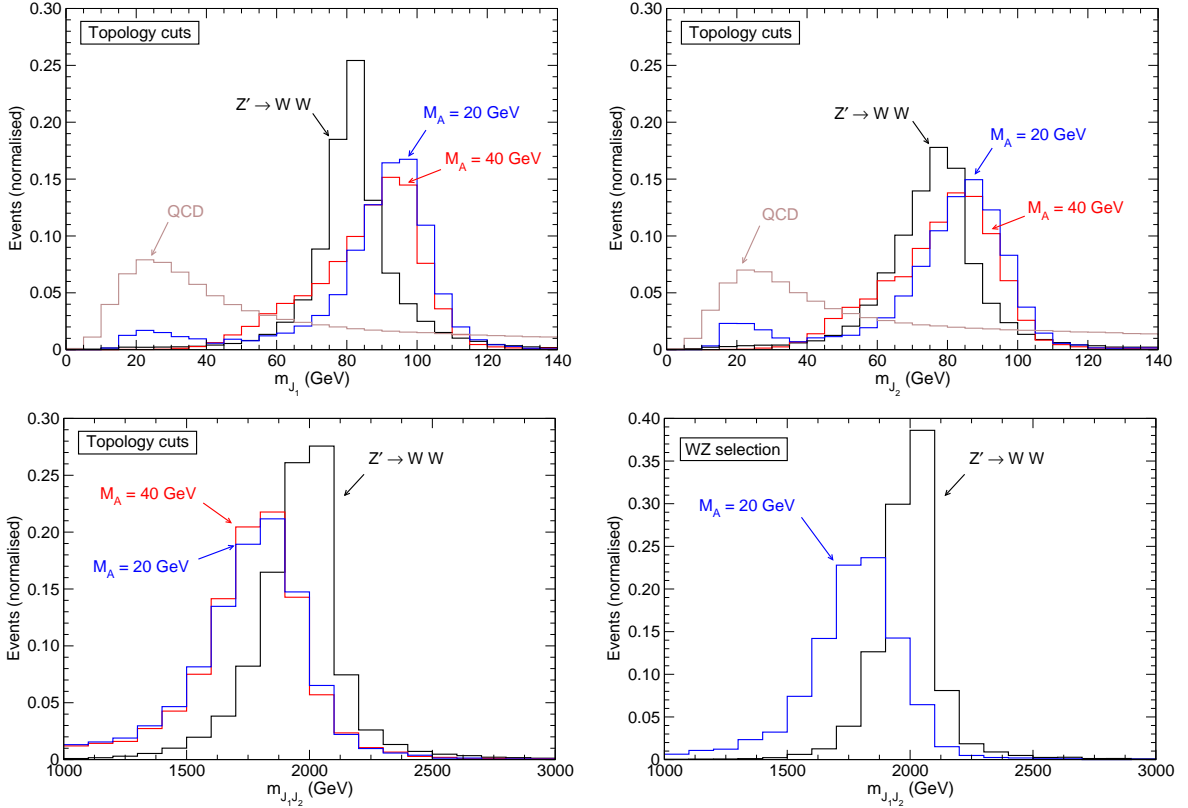

\begin{center}
\begin{tabular}{cc}
\includegraphics[height=5.25cm,clip=]{Figs/mJ1_a-Zp.eps} &
\includegraphics[height=5.25cm,clip=]{Figs/mJ2_a-Zp.eps} \\
\includegraphics[height=5.25cm,clip=]{Figs/mJJ_a-Zp.eps}  &
\includegraphics[height=5.25cm,clip=]{Figs/mJJ_at-Zp.eps}
\end{tabular}
\caption{Jet and dijet mass distributions for $R \to SS$ and $Z' \to W W$ signals and the QCD dijet background for the ATLAS analysis}
\label{fig:XX-ATLAS2}
\end{center}
\end{figure}

The trimming algorithm used by the ATLAS Collaboration allows for a slightly better jet mass determination for stealth bosons than in the CMS analyses. We show in Fig.~\ref{fig:XX-ATLAS2} (top) the masses of the leading and subleading jet after the topological selection. The shapes are less peaked than for $W$ bosons. The dijet mass distributions on the bottom panels are very wide, and similar after topology cuts only (left) and with the full $WZ$ tagging (right). In the latter case the corresponding plot for $M_A = 40$ GeV is not shown because the signal is tiny.

To conclude this section, we compare in table~\ref{tab:eff-XX} the efficiencies for the $WZ$ selection in the ATLAS Run 1~\cite{Aad:2015owa} and Run 2~\cite{ATLAS:2016yqq} searches. The results are indeed eloquent.
\begin{table}[htb]
\begin{center}
\begin{tabular}{lcc}
& Run 1 & Run 2  \\
$WW$ & 0.14 & 0.16  \\
QCD & $2.6 \times 10^{-4}$ & $2.0 \times 10^{-5}$  \\
$M_A = 20~\text{GeV}$ & 0.10 & 0.065  \\
$M_A = 40~\text{GeV}$ & $3.6 \times 10^{-3}$ & $2.4 \times 10^{-3}$
\end{tabular}
\caption{Efficiencies of $R \to SS$ and $Z' \to W W$ signals and the QCD dijet background for the $WZ$ selection in ATLAS hadronic diboson searches.}
\label{tab:eff-XX}
\end{center}
\end{table}
While for a true diboson signal the efficiency of the ATLAS Run 2 event selection is larger than in Run 1, for the $R \to SS$ signals it is noticeably smaller --- which might have motivated the disappearance of the excess in the first Run 2 result with the nominal selection, if it were due to a signal of this type. Of course, the precise numbers vary with $M_A$, but the trend is the correct one. The comparison of these figures with the efficiency for the CMS selection is not meaningful because for CMS analyses the jet mass windows are narrower, and the events in the $WW$, $WZ$ and $WW$ categories, which are disjoint, are combined to obtain the limits on a potential signal. Besides, we point out that, as argued in the introduction from a naive parton luminosity scaling, the efficiency for the background is reduced by around an order of magnitude in the Run 2 selection with respect to Run 1.


\section{Diboson-like $R \to VS$ decays}
\label{sec:3}

\begin{figure}[t]
\begin{center}
\includegraphics[height=3cm,clip=]{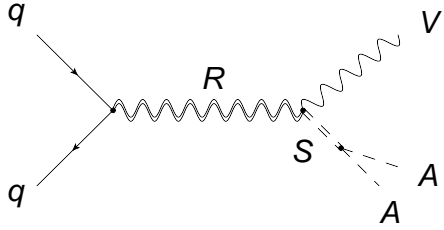}
\caption{Sample diagram for a $R \to VS $ signal.}
\label{fig:diag2}
\end{center}
\end{figure}

We repeat the same procedure of the previous section for a $R \to VS$ decay such as the one depicted in Fig.~\ref{fig:diag2}, which yields a triboson final state but with a diboson-like topology.
Such a signal can originate for example in left-right models from the decay $W' \to W H_1^0$, with subsequent $H_1^0 \to A^0 A^0$ decay~\cite{Aguilar-Saavedra:2015iew}, or also from $Z' \to Z H_1^0$. We consider the former process, with $S = H_1^0$, $A=A^0$ and the same parameters as in section~\ref{sec:2}. The results, for what concerns the hadronic diboson searches, would be analogous for a signal of the latter type with a $Z$ boson in the final state.

\begin{figure}[t]
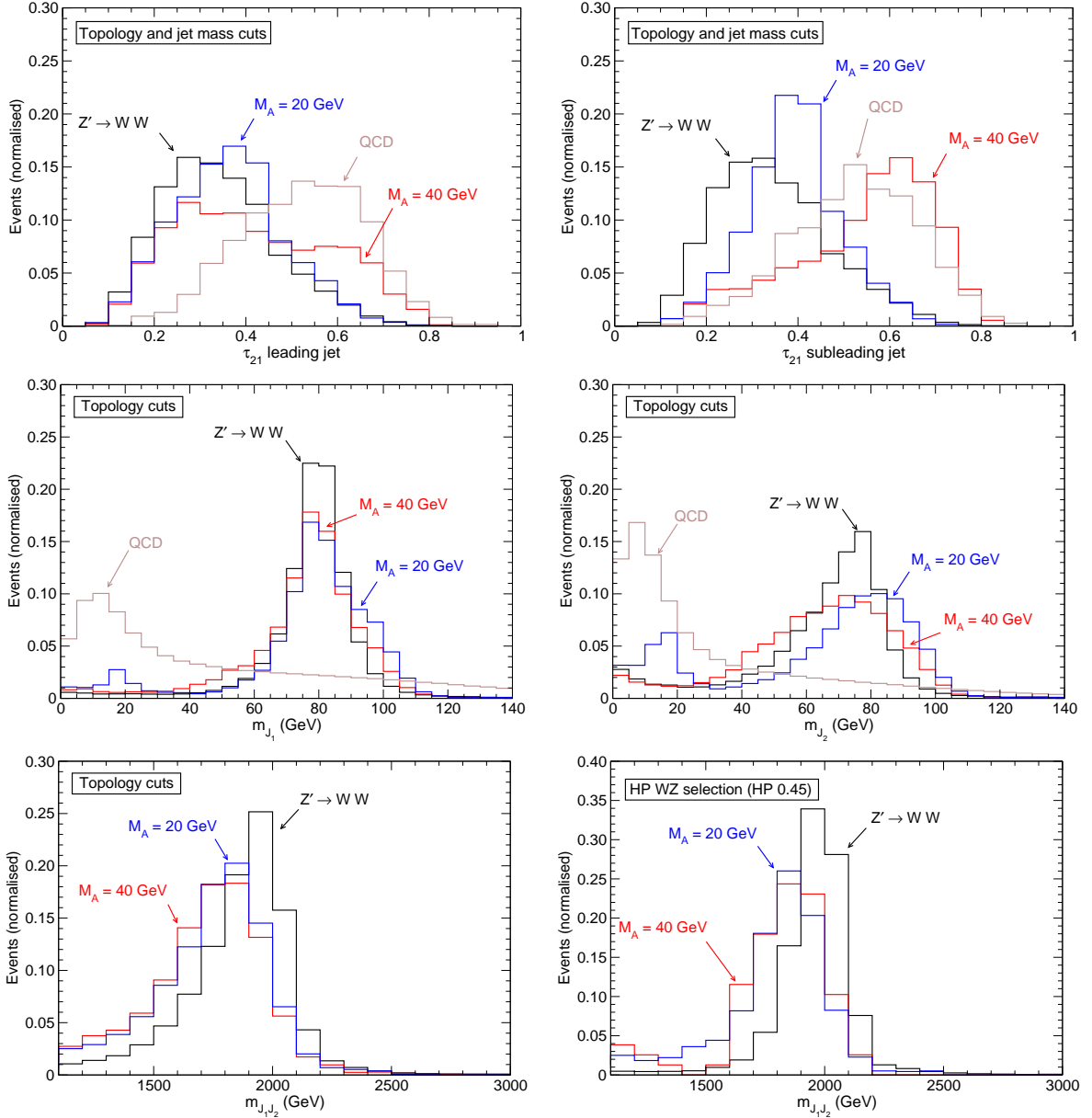

\begin{center}
\begin{tabular}{cc}
\includegraphics[height=5.25cm,clip=]{Figs/tau21L-WpD.eps} &
\includegraphics[height=5.25cm,clip=]{Figs/tau21S-WpD.eps} \\
\includegraphics[height=5.25cm,clip=]{Figs/mJ1_c-WpD.eps} &
\includegraphics[height=5.25cm,clip=]{Figs/mJ2_c-WpD.eps} \\
\includegraphics[height=5.25cm,clip=]{Figs/mJJ_c-WpD.eps} &
\includegraphics[height=5.25cm,clip=]{Figs/mJJ_ct-WpD.eps}
\end{tabular}
\caption{Distributions for the $R \to VS$ and $Z' \to W W$ signals and the QCD dijet background in the CMS analysis. Top: $N$-subjettiness ratio $\tau_{21}$; middle: jet mass distributions; bottom: dijet mass distributions.}
\label{fig:VX-CMS}
\end{center}
\end{figure}

For the CMS analysis the $\tau_{21}$ distributions for the leading and subleading jets are separately shown in Fig.~\ref{fig:VX-CMS}. These distributions are a combination of the ones for the $W$ and $S$ bosons in Fig.~\ref{fig:XX-CMS1}, but they are obviously correlated: always one of the two jets, the one corresponding to $S$, has a large $\tau_{21}$. For this reason, it is illustrative to quantify the efficiency drop when one changes the $\tau_{21}$ cut from 0.45 to 0.35,
\begin{align*}
& M_A = 20~\text{GeV}: && 0.068 \to 0.014 && ( \times \; 0.2) \,, \\
& M_A = 40~\text{GeV}: && 3.9 \times 10^{-3} \to 3.5 \times 10^{-4} && ( \times \; 0.09) \,.
\end{align*}
The masses of the leading and subleading jets are also shown in Fig.~\ref{fig:VX-CMS}. The distributions are narrower than for a $R \to SS$ decay, as it may be expected because one of the decay products is a massive SM boson. The dijet invariant mass distributions are wide, peaking at invariant masses slightly lower than the resonance mass. The distributions after boson tagging are slightly sharper, comparable to a true diboson resonance.

\begin{figure}[p]
\begin{center}
\begin{tabular}{cc}
\includegraphics[height=5.25cm,clip=]{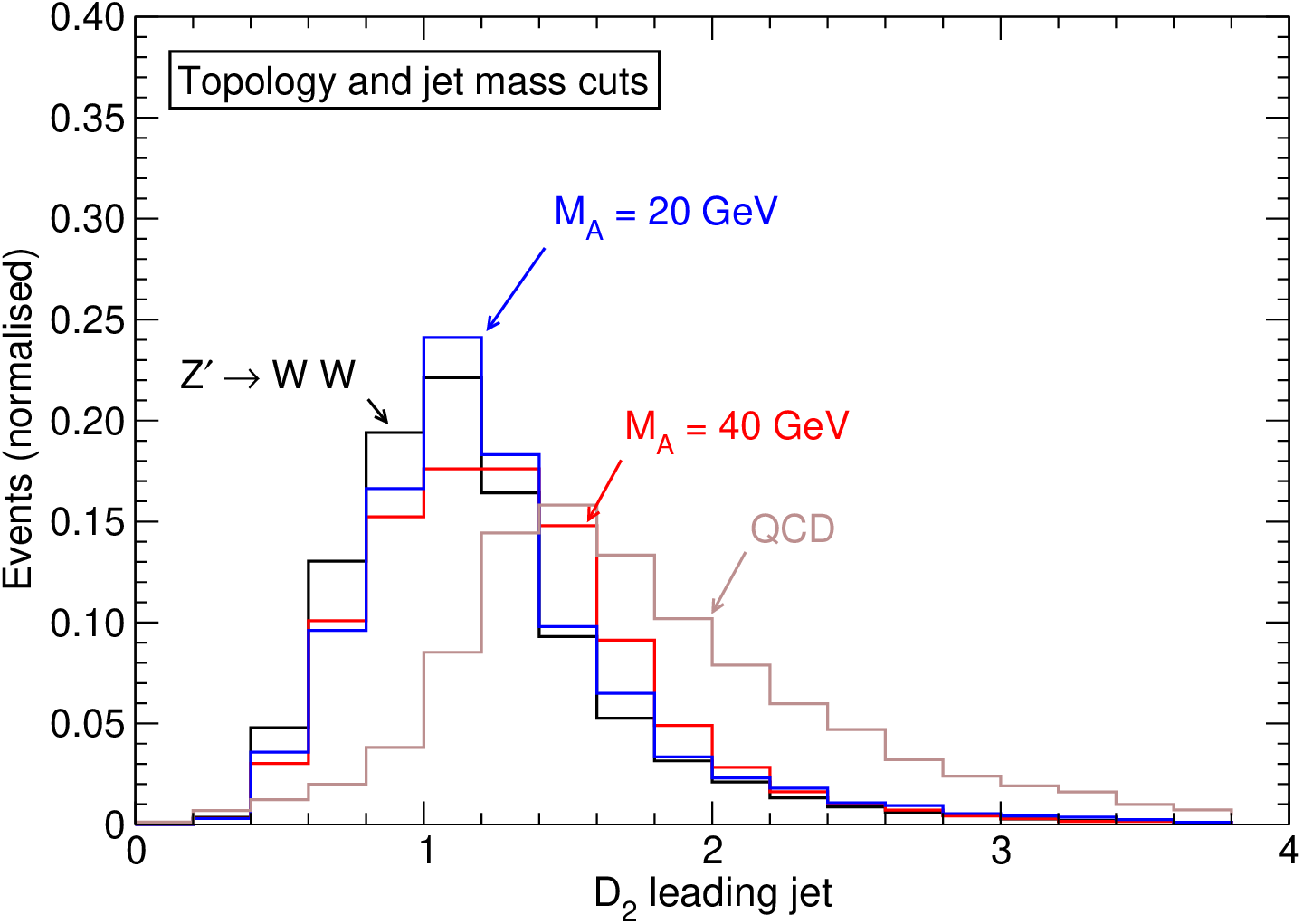} &
\includegraphics[height=5.25cm,clip=]{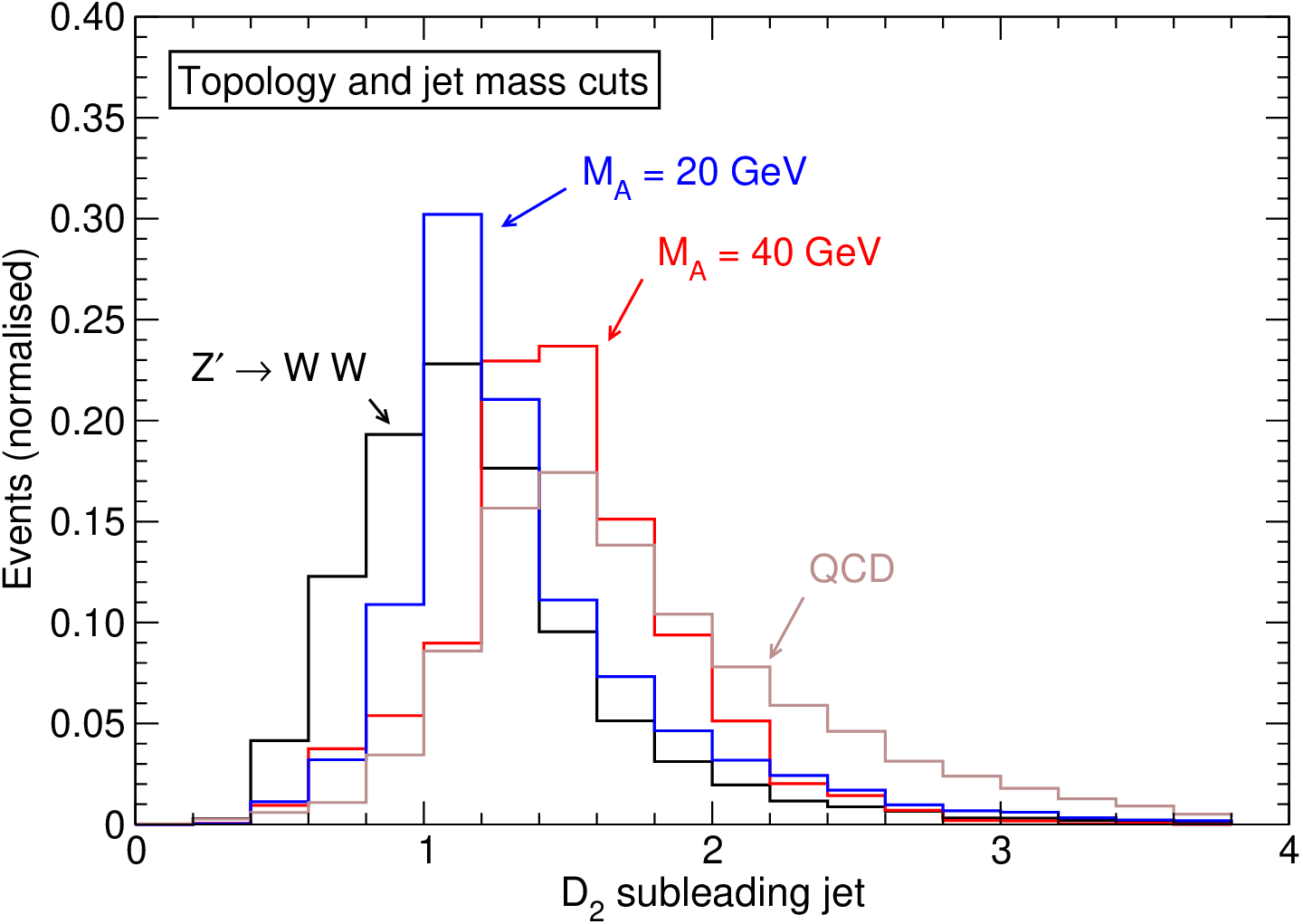} \\
\includegraphics[height=5.25cm,clip=]{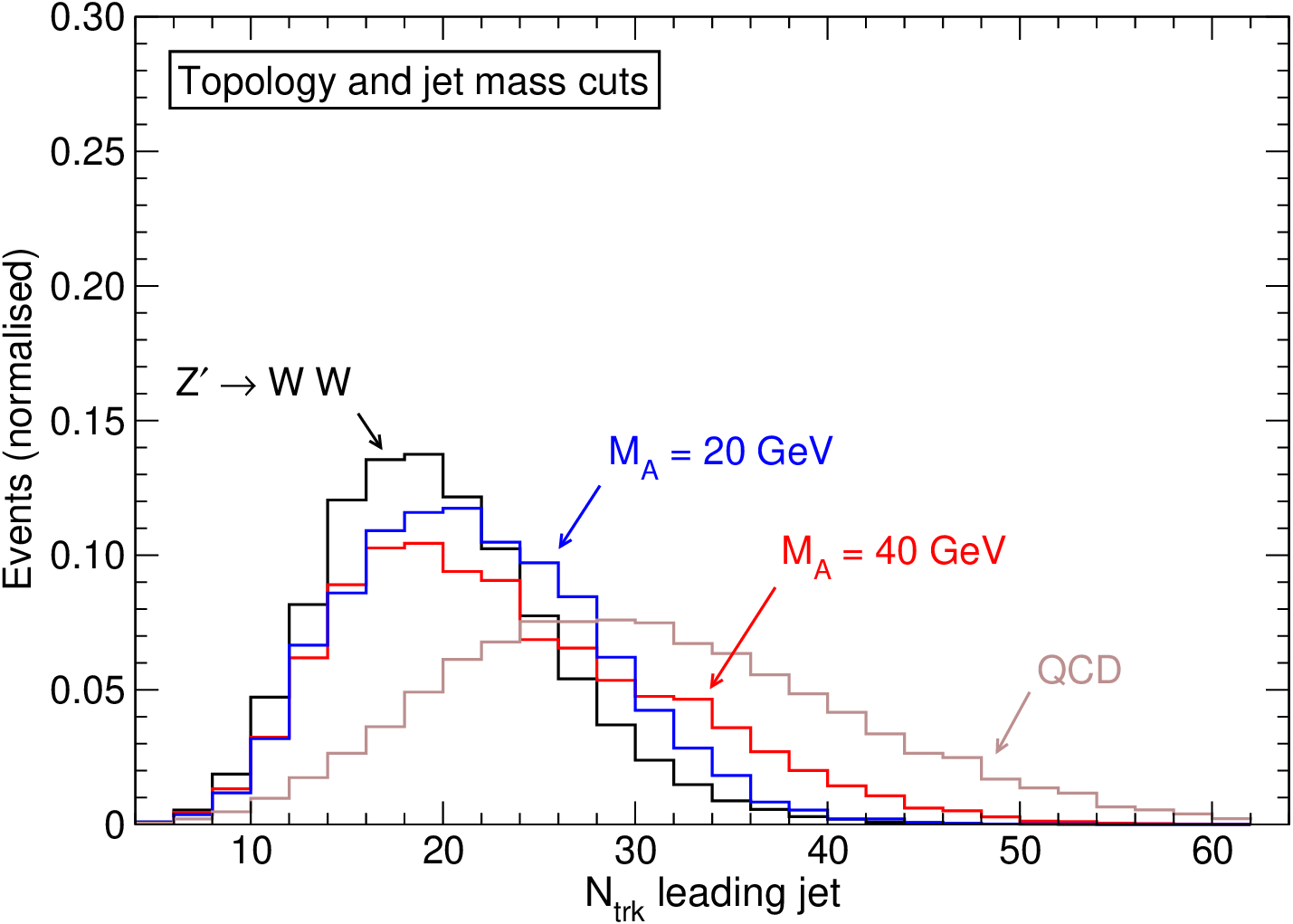} &
\includegraphics[height=5.25cm,clip=]{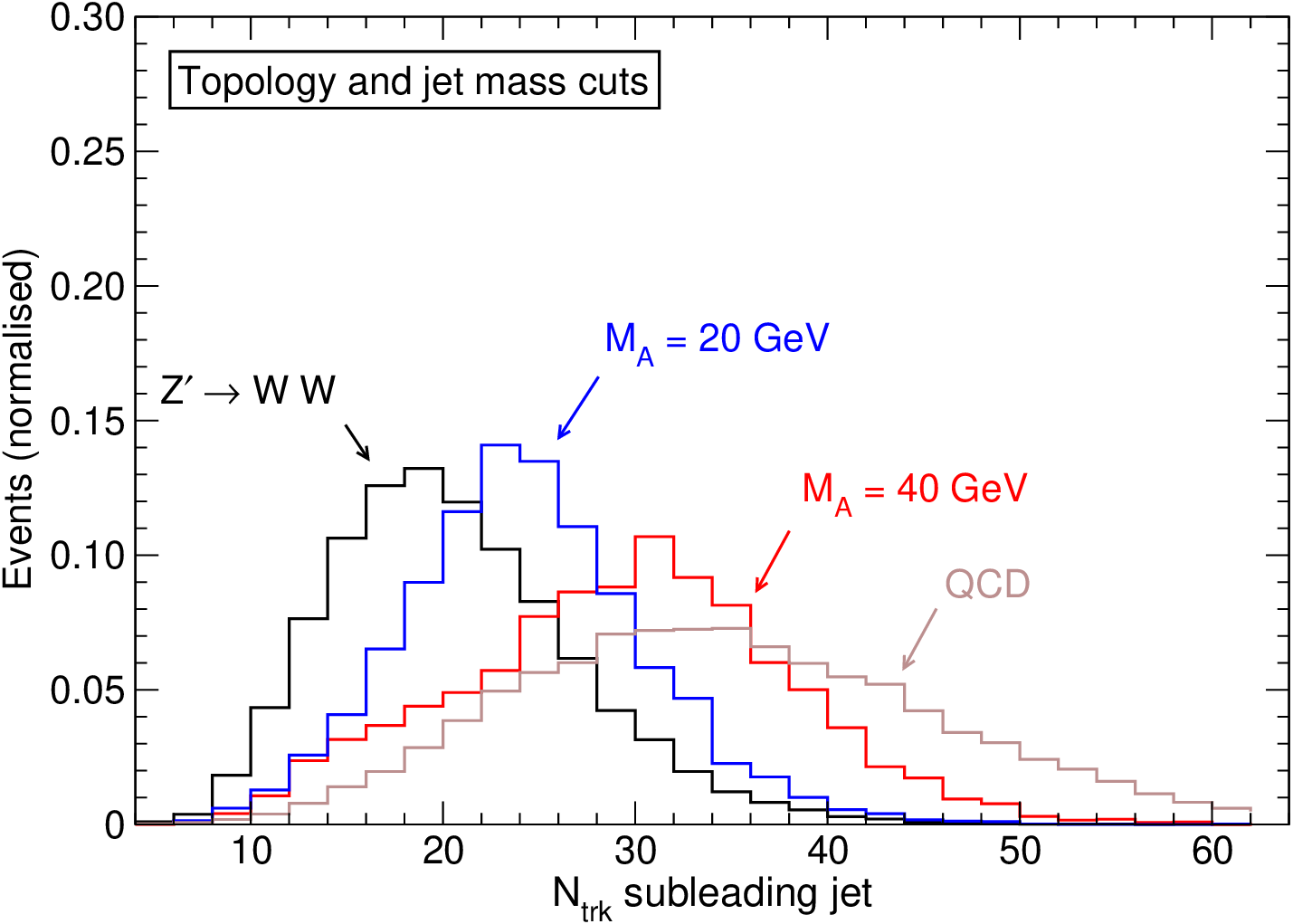} \\
\includegraphics[height=5.25cm,clip=]{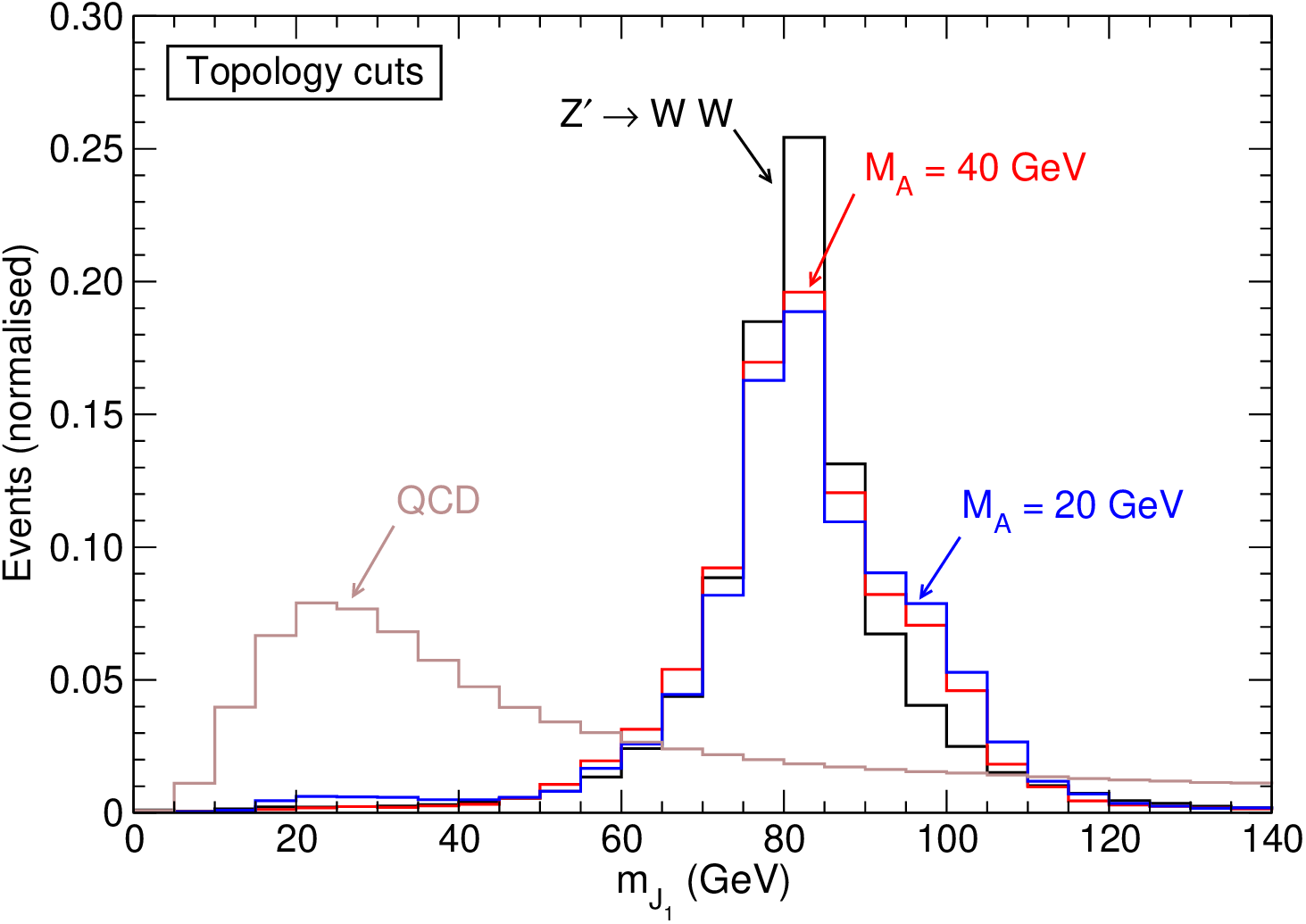} &
\includegraphics[height=5.25cm,clip=]{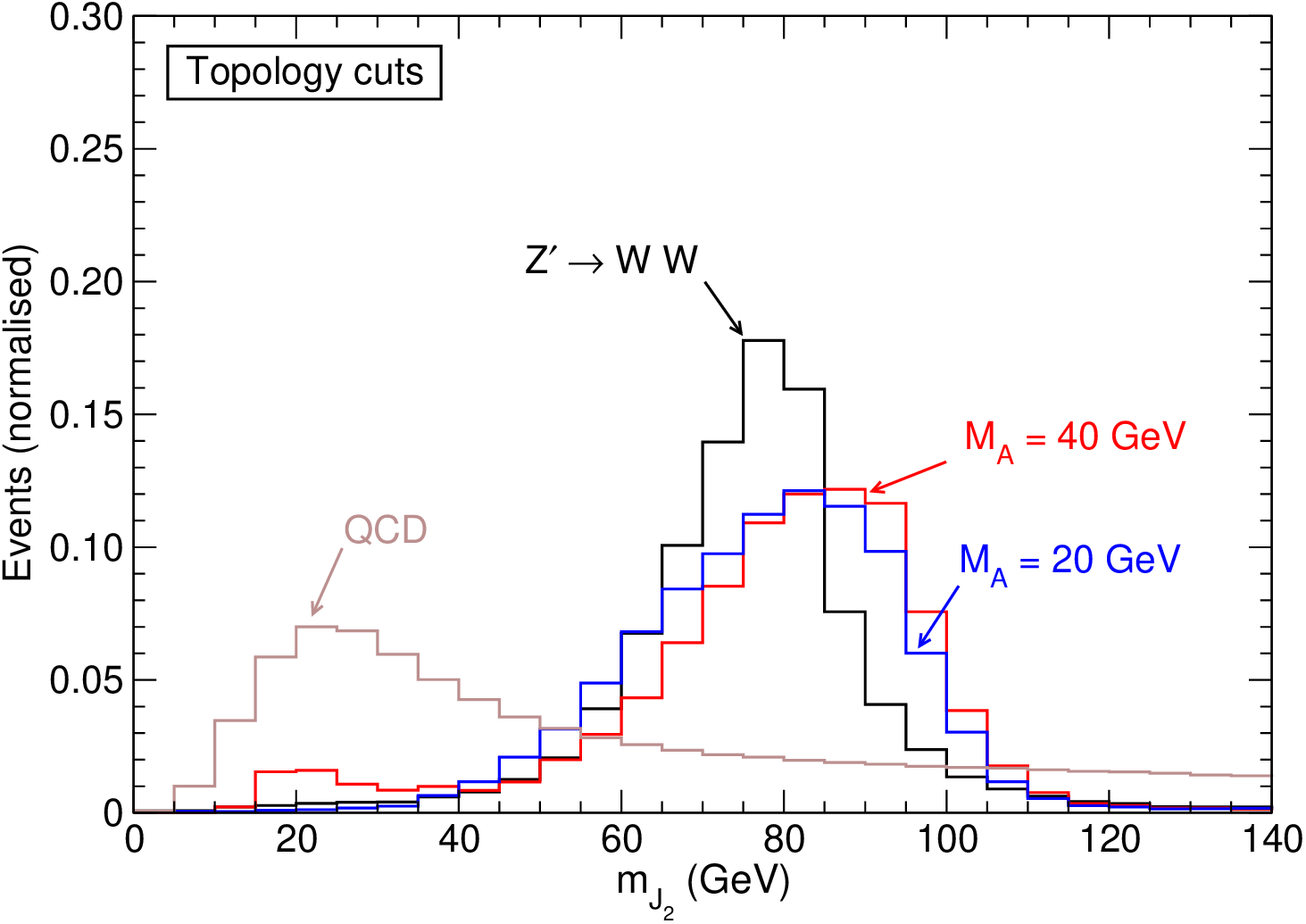} \\
\includegraphics[height=5.25cm,clip=]{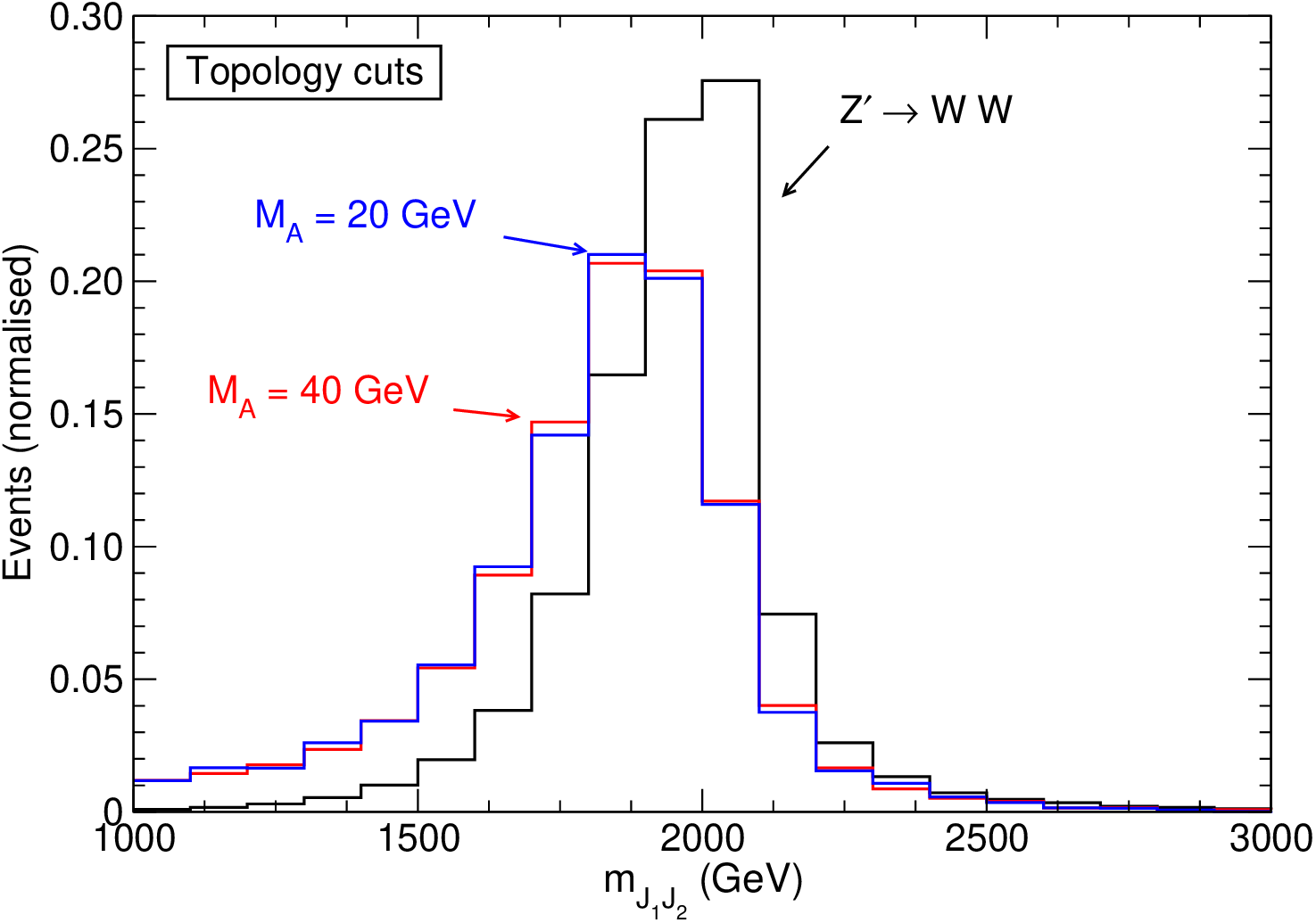}  &
\includegraphics[height=5.25cm,clip=]{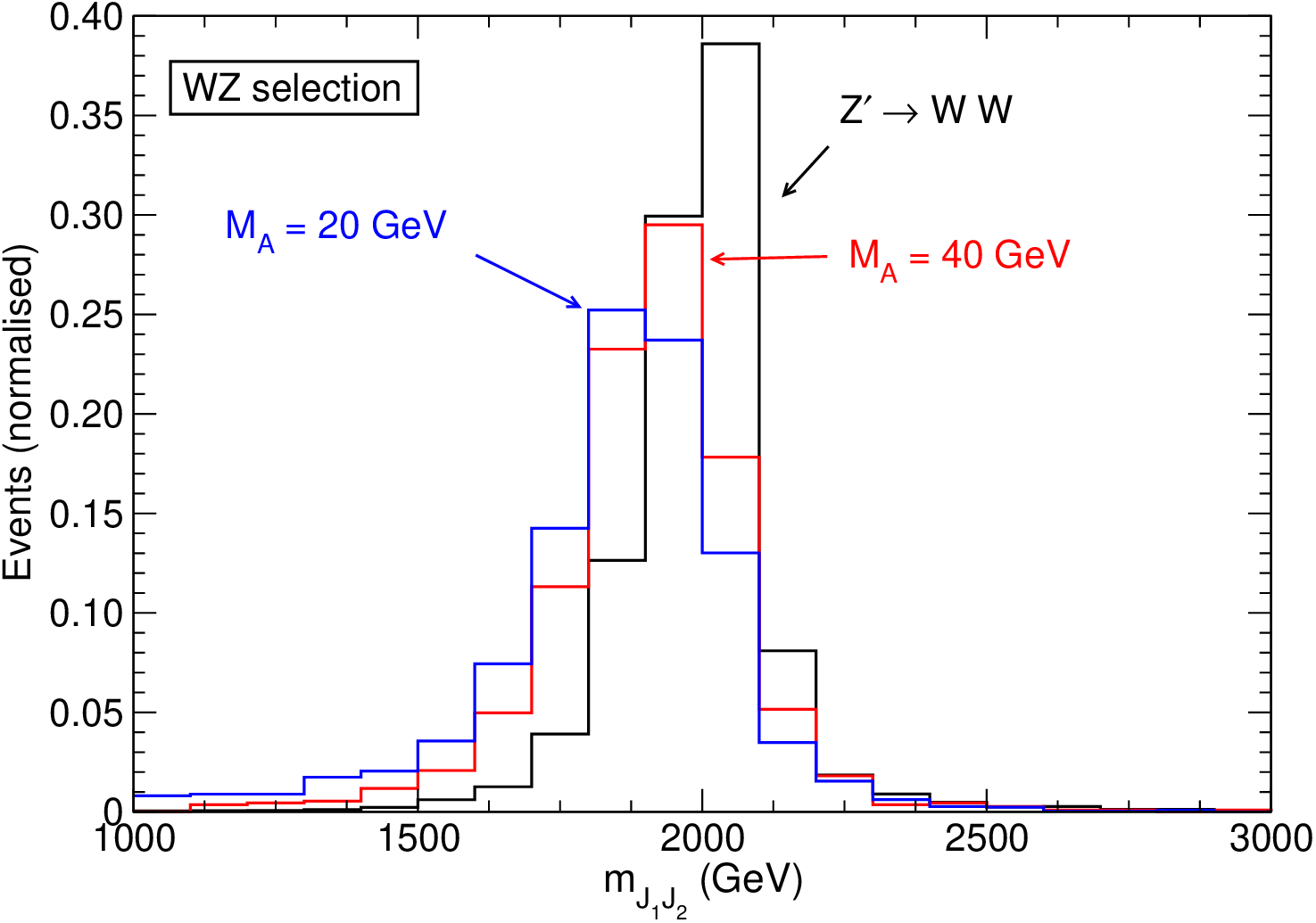}
\end{tabular}
\caption{Distributions for the $R \to VS$ and $Z' \to W W$ signals and the QCD dijet background in the ATLAS analysis. Top: $D_2$ variable; second row: number of tracks; third row: jet mass; bottom: dijet invariant mass.}
\label{fig:VX-ATLAS}
\end{center}
\end{figure}

For the ATLAS Run 2 analyses the $D_2$ and $\ntrk$ boson tagging variables are shown in Fig.~\ref{fig:VX-ATLAS}.
For the leading jet the distributions are similar for the three signals, indicating that the leading jet in $R \to VS$ is the SM vector boson in most cases. The $D_2$ tagging requirement on both jets has an efficiency of 0.35 and 0.18 for $M_A = 20$ and 40 GeV, respectively, larger than for the $R \to SS$ signal. Here it is also interesting to see the effect of the $\ntrk < 30$ cut, which reduces the signal efficiencies for the $WZ$ selection as follows,
\begin{align*}
& M_A = 20~\text{GeV}: && 0.13 \to 0.096 && ( \times \; 0.74) \,, \\
& M_A = 40~\text{GeV}: && 0.055 \to 0.020  && ( \times \; 0.36) \,.
\end{align*}
For this type of signals an upper cut on the number of tracks may be counterproductive too, especially for heavier $A$.

Besides the jet substructure differences found, the topology and jet mass of the $R \to VS$ signal are not very different from a true diboson resonance, as it can be seen by comparing the rest of distributions presented in Fig.~\ref{fig:VX-ATLAS}, namely the masses of the leading and subleading jet after the topological selection, the dijet invariant mass after topological selection and also with the final $WZ$ tagging. The distributions are less peaked than for a true diboson but still they are alike. Finally, we collect in table~\ref{tab:eff-VX} the efficiencies for the full $WZ$ tagging in the ATLAS Run 1 / Run 2 analyses. At variance with the results shown in the previous section, the efficiencies for both ATLAS analyses are comparable, but slightly smaller in Run 2. Such a signal might also accommodate the size of the excesses observed in the different searches.

\begin{table}[htb]
\begin{center}
\begin{tabular}{lcc}
& Run 1 &  Run 2  \\
$M_A = 20~\text{GeV}$ & 0.11 & 0.096 \\
$M_A = 40~\text{GeV}$ & 0.022 & 0.020 
\end{tabular}
\caption{Efficiencies of $R \to VS$ signals for the $WZ$ selection in ATLAS hadronic diboson searches.}
\label{tab:eff-VX}
\end{center}
\end{table}


\section{Triboson-like $R \to VVS$ decays}
\label{sec:4}

\begin{figure}[htb]
\begin{center}
\includegraphics[height=3cm,clip=]{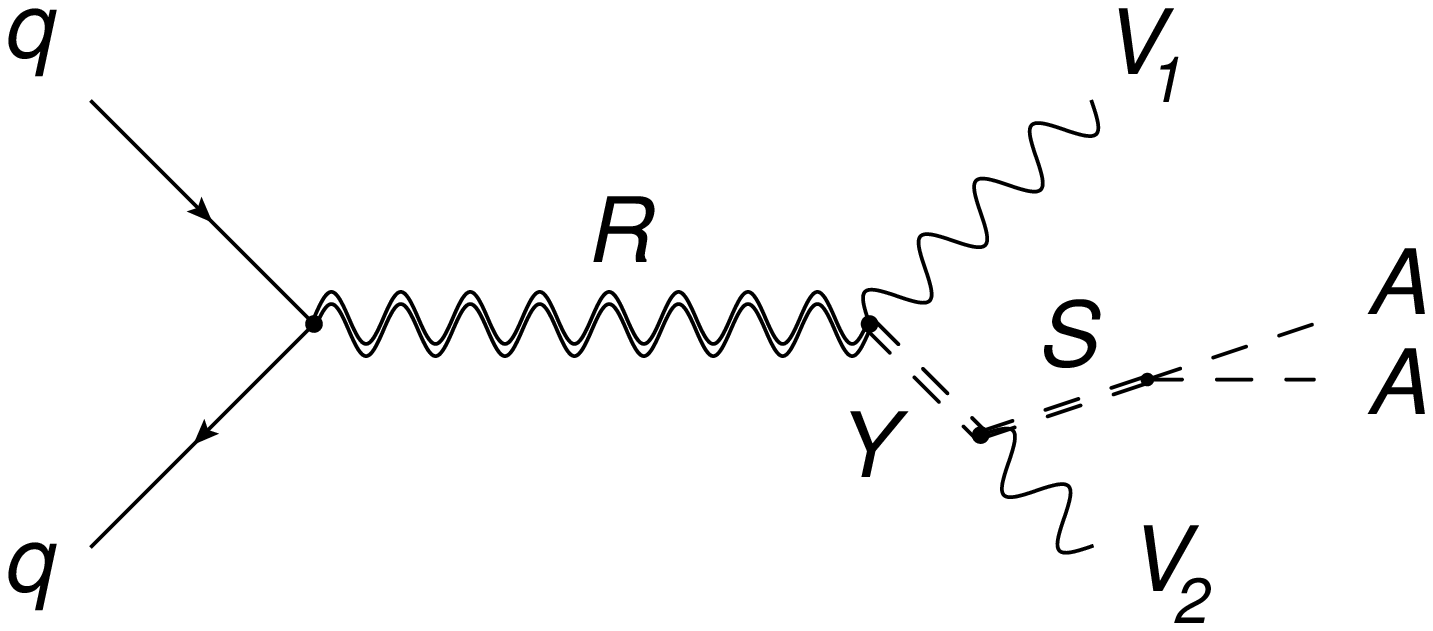}
\caption{Sample diagram for a $R \to V_1 Y \to V_1 V_2 S $ signal.}
\label{fig:diag3}
\end{center}
\end{figure}

As our third example we consider a cascade decay $R \to VY \to VVS$, with subsequent decay $S \to AA$, as depicted in Fig.~\ref{fig:diag3}. This is a quadriboson signal but with a triboson-like topology. 
There are two crucial differences with the previously seen resonance decays. First, there are already two SM bosons in the final state, which produce fat jets with a higher boson tagging efficiency. Second, the invariant mass of the two selected jets does not concentrate at a maximum near or below the resonance mass, but instead the dijet invariant mass distributions are broader\cite{Aguilar-Saavedra:2015rna}. An example of this cascade decay chain is $W' \to Z H^\pm \to Z W^\pm H_1^0$ in left-right models~\cite{Aguilar-Saavedra:2015iew}. This is the signal used in our simulations, taking $Y = H^\pm$ with $M_Y = 500$ GeV, $S = H_1^0$, $A = A^0 \to b \bar b$.

\begin{figure}[t]
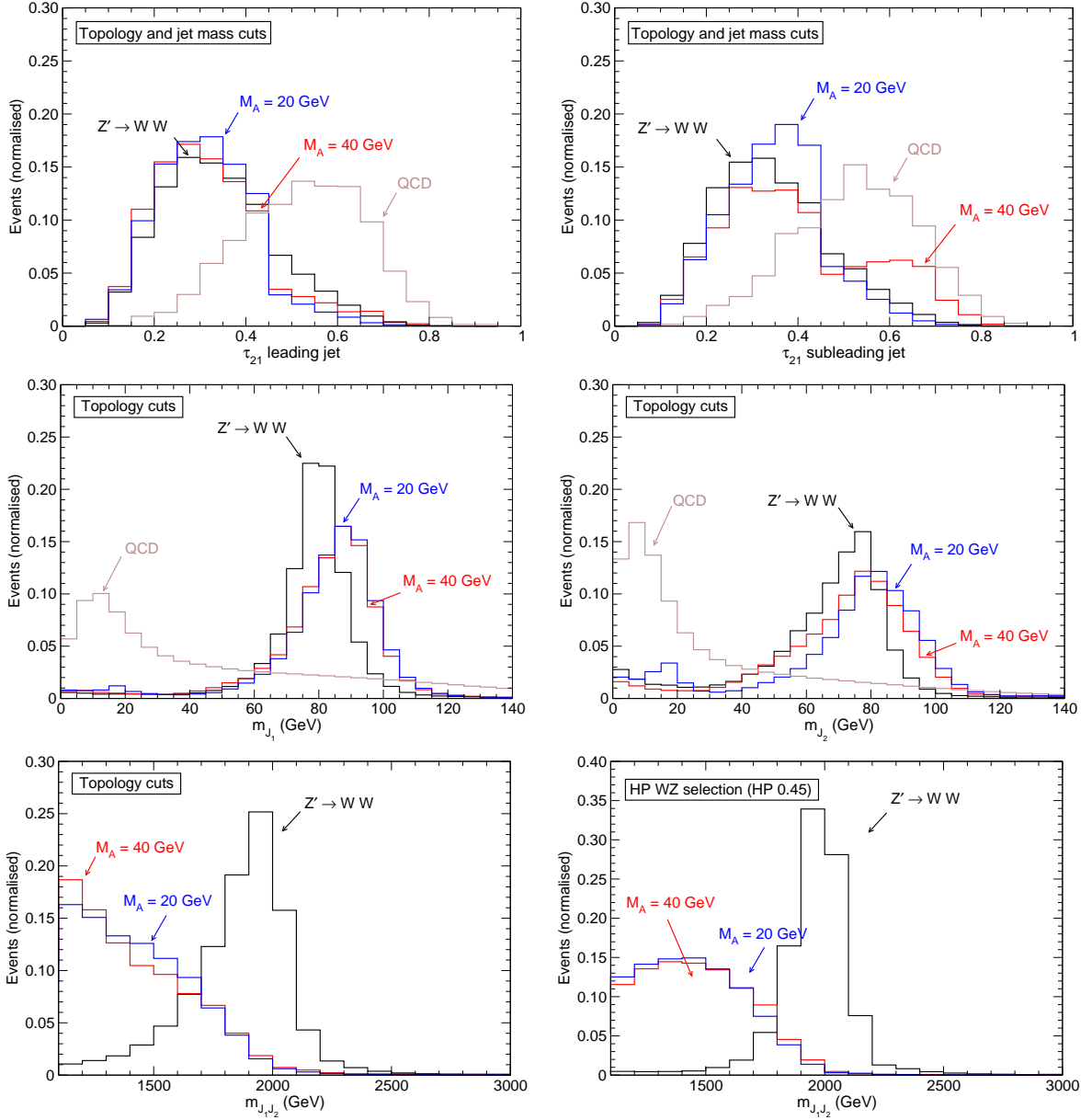

\begin{center}
\begin{tabular}{cc}
\includegraphics[height=5.25cm,clip=]{Figs/tau21L-WpT.eps} &
\includegraphics[height=5.25cm,clip=]{Figs/tau21S-WpT.eps}  \\
\includegraphics[height=5.25cm,clip=]{Figs/mJ1_c-WpT.eps} &
\includegraphics[height=5.25cm,clip=]{Figs/mJ2_c-WpT.eps} \\
\includegraphics[height=5.25cm,clip=]{Figs/mJJ_c-WpT.eps} &
\includegraphics[height=5.25cm,clip=]{Figs/mJJ_ct-WpT.eps}
\end{tabular}
\caption{Distributions for the $R \to VVS$ and $Z' \to W W$ signals and the QCD dijet background in the CMS analysis. Top: $N$-subjettiness ratio $\tau_{21}$; middle: jet mass distributions; bottom: dijet mass distributions.}
\label{fig:VVX-CMS}
\end{center}
\end{figure}

For this signal we collect the relevant distributions for the CMS analyses in Fig.~\ref{fig:VVX-CMS}. 
The $\tau_{21}$ distribution for the leading jet is quite close to the one for a $Z' \to W W$ diboson signal, while there are some differences for the subleading jet. (This indicates that the leading jet is the $Z$ boson from the heavy resonance decay in most cases.) The efficiency for $\tau_{21}$ and mass tagging does not change as dramatically as for the previous $R \to SS$ and $R \to VS$ examples,
\begin{align*}
& M_A = 20~\text{GeV}: && 0.13 \to 0.053 && ( \times \; 0.42) \,, \\
& M_A = 40~\text{GeV}: && 0.082 \to 0.039 && ( \times \; 0.47) \,,
\end{align*}
that is, it decreases by nearly the same amount as for a $Z' \to WW$ signal. The jet mass distributions are not very different from a diboson resonance either. Note that here the leading jet mass concentrates around $M_Z$, as corresponds to the simulated signal. Sharp differences are found in the dijet invariant mass distributions in the bottom panels of Fig.~\ref{fig:VVX-CMS}. The $m_{JJ}$ distribution does not display any resonance-like structure before the boson tagging (left) and a wide bump, similar to the ones found for triboson signals in Ref.~\cite{Aguilar-Saavedra:2016xuc}, appears after boson tagging.

\begin{figure}[p]
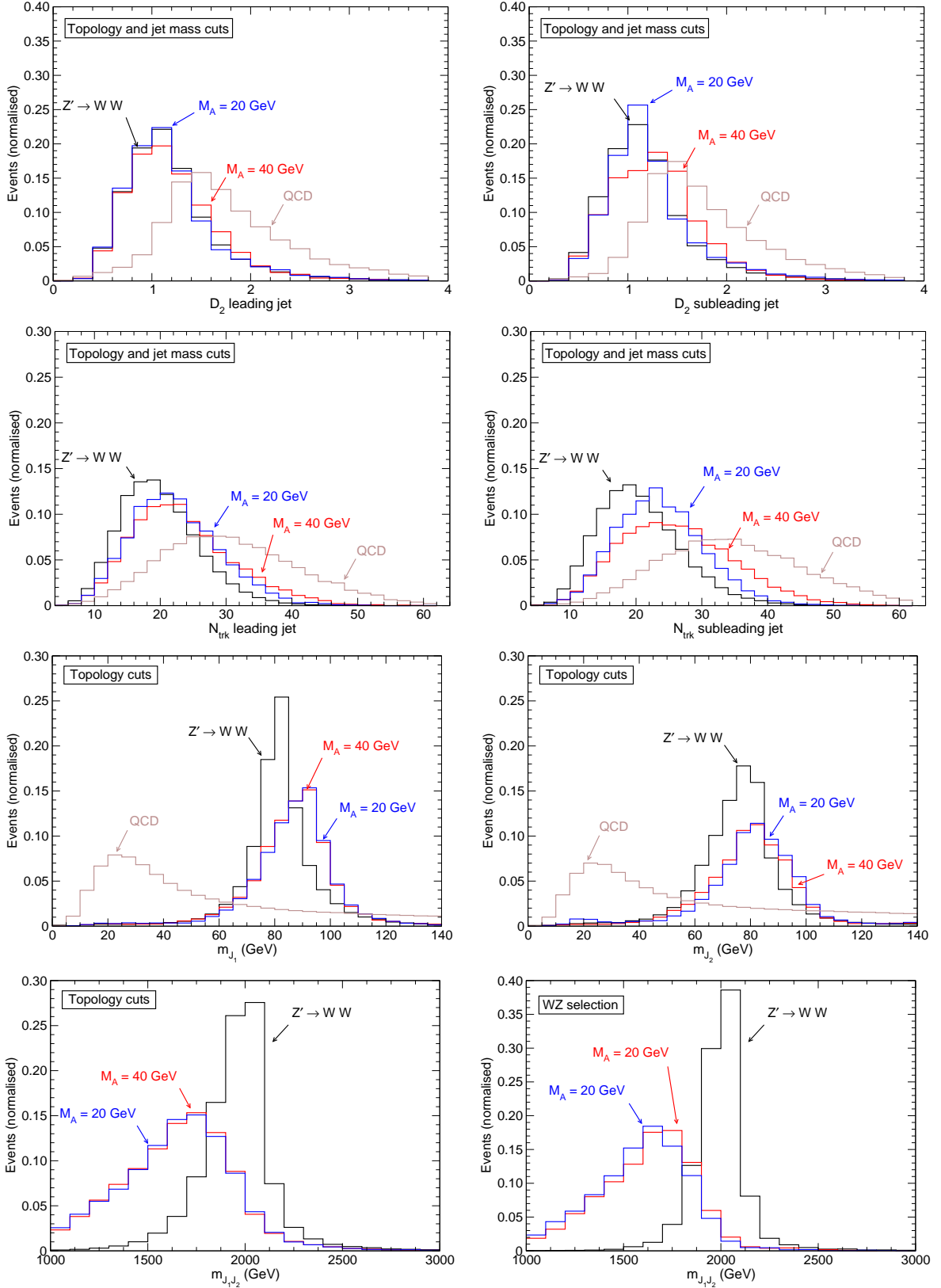

\begin{center}
\begin{tabular}{cc}
\includegraphics[height=5.25cm,clip=]{Figs/D2L-WpT.eps} &
\includegraphics[height=5.25cm,clip=]{Figs/D2S-WpT.eps} \\
\includegraphics[height=5.25cm,clip=]{Figs/NtrkL-WpT.eps} &
\includegraphics[height=5.25cm,clip=]{Figs/NtrkS-WpT.eps} \\
\includegraphics[height=5.25cm,clip=]{Figs/mJ1_a-WpT.eps} &
\includegraphics[height=5.25cm,clip=]{Figs/mJ2_a-WpT.eps} \\
\includegraphics[height=5.25cm,clip=]{Figs/mJJ_a-WpT.eps}  &
\includegraphics[height=5.25cm,clip=]{Figs/mJJ_at-WpT.eps}
\end{tabular}
\caption{Distributions for the $R \to VVS$ and $Z' \to W W$ signals and the QCD dijet background in the ATLAS analysis. Top: $D_2$ variable; second row: number of tracks; third row: jet mass; bottom: dijet invariant mass.}
\label{fig:VVX-ATLAS}
\end{center}
\end{figure}

The distributions for the ATLAS Run 2 analyses are given in Fig.~\ref{fig:VVX-ATLAS}.
The $D_2$ variable for the leading and subleading jets in $R \to VVS$ in the top panels are similar to the ones for the benchmark $Z' \to WW$ signal, and the effiency for the $D_2$ cut in both jets is 0.4 for $M_A = 20$ GeV and 0.33 for $M_A = 40$ GeV, close to the value of 0.5 for a $WW$ signal. The number of tracks per jet  is slightly higher, as it can be seen from the plots in the second row. For the $WZ$ selection, the $\ntrk < 30$ cut reduces the signal efficiencies as follows,
\begin{align*}
& M_A = 20~\text{GeV}: && 0.082 \to 0.063 && ( \times \; 0.77) \,, \\
& M_A = 40~\text{GeV}: && 0.058 \to 0.042  && ( \times \; 0.72) \,.
\end{align*}
That is, the signal suppresion is higher than for a true diboson signal but not dramatic. The jet mass distributions are relatively narrow, and the dijet invariant mass distributions do not display a peak, even after the application of the full $WZ$ selection. This contrasts with the behaviour of true triboson signals, for which the invariant mass distribution of the two leading jets is shaped to a peak by the application of topological and boson tagging cuts~\cite{Aguilar-Saavedra:2015rna,Aguilar-Saavedra:2016xuc}. The comparison of the efficiencies for ATLAS Run 1 and Run 2 analyses is given in table~\ref{tab:eff-VVX}. They are slightly smaller in Run 2.
Overall, we find that except for some efficiency decrease when the boson tagging requirements are tightened, the $VVS$ signals behave in much the same way as the triboson signals studied before~\cite{Aguilar-Saavedra:2016xuc}.

\begin{table}[htb]
\begin{center}
\begin{tabular}{lcc}
& Run 1 &  Run 2  \\
$M_A = 20~\text{GeV}$ & 0.074 & 0.063 \\
$M_A = 40~\text{GeV}$ & 0.050 & 0.042 
\end{tabular}
\caption{Efficiencies of $R \to VS$ signals for the $WZ$ selection in ATLAS hadronic diboson searches.}
\label{tab:eff-VVX}
\end{center}
\end{table}


\section{Discussion}
\label{sec:5}

New physics interpretations of the ATLAS Run 1 diboson excess~\cite{Aad:2015owa} can explain the absence of signals in the semileptonic channels, either because the signal is due to a $VVX$ triboson resonance~\cite{Aguilar-Saavedra:2015rna} or by assuming a particle with a mass close to $M_W,M_Z$ and decaying into $q \bar q$ only~\cite{Allanach:2015blv}. (Interpretations of this excess as a diboson resonance were early disfavoured, and have been already excluded with the first Run 2 results.) Nevertheless, the main difficulty for these proposals is posed by the most recent ATLAS and CMS searches in the hadronic channels, some of which again exhibit some mild excesses at the same mass, but with sizes that are apparently inconsistent. Even taking into account that, as shown in Ref.~\cite{Aguilar-Saavedra:2017iso}, part of the excess may be absorbed by the normalisation of the background if statistics are small, one would expect more significant excesses in the large samples collected in Run 2.

The `stealth bosons' $S$ introduced in this paper, which decay into four quarks that merge into a single fat jet, can succesfully address apparent inconsistencies among mild excesses in different searches like those. For stealth bosons decaying $S \to AA$ we have found that:
\begin{enumerate}
\item For the $R \to SS$ and $R \to VS$ signals the boson tagging efficiency for CMS analyses drops when the cut on the $\tau_{21}$ variable, used to measure the two-pronged structure of the jets, is tightened. This may explain why the latest CMS analysis with 35.9 fb$^{-1}$~\cite{Sirunyan:2017acf} does not observe a much larger excess, and the previous one with 12.9 fb$^{-1}$~\cite{CMS:2016mwi} had a small deviation at the $1\sigma$ level.
\item For $R \to SS$ especially, and to some extent for $R \to VS$ and $R \to VVS$ too, the efficiency for the ATLAS Run 2 event selection is smaller than at Run 1, explaining why the Run 2 excess with 15.5 fb$^{-1}$~\cite{ATLAS:2016yqq} was of only $2\sigma$, and with 3.2 fb$^{-1}$ the deviation was at the $1\sigma$ level~\cite{ATLAS:2015msj}.
\item The upper cut on the number of tracks $\ntrk$ applied by the ATLAS Collaboration washes out a $R \to SS$ signal. The suppresion is milder for $R \to VS$ and $R \to VVS$. This might explain why the dijet mass distribution without this cut exhibits a bump already with 3.2 fb$^{-1}$~\cite{ATLAS:2015msj}, as already pointed out~\cite{Aguilar-Saavedra:2015iew,Aguilar-Saavedra:2016xuc}.
\end{enumerate} 
These results are quite independent of the heavy resonance mass, as long as the decay products are sufficiently boosted. This already happens for $M_R \geq 1.5$ TeV. And the results are applicable, at least qualitatively, to heavier stealth bosons and the decay $S \to ZA$, as seen in appendix~\ref{sec:b}.

\begin{figure}[t]
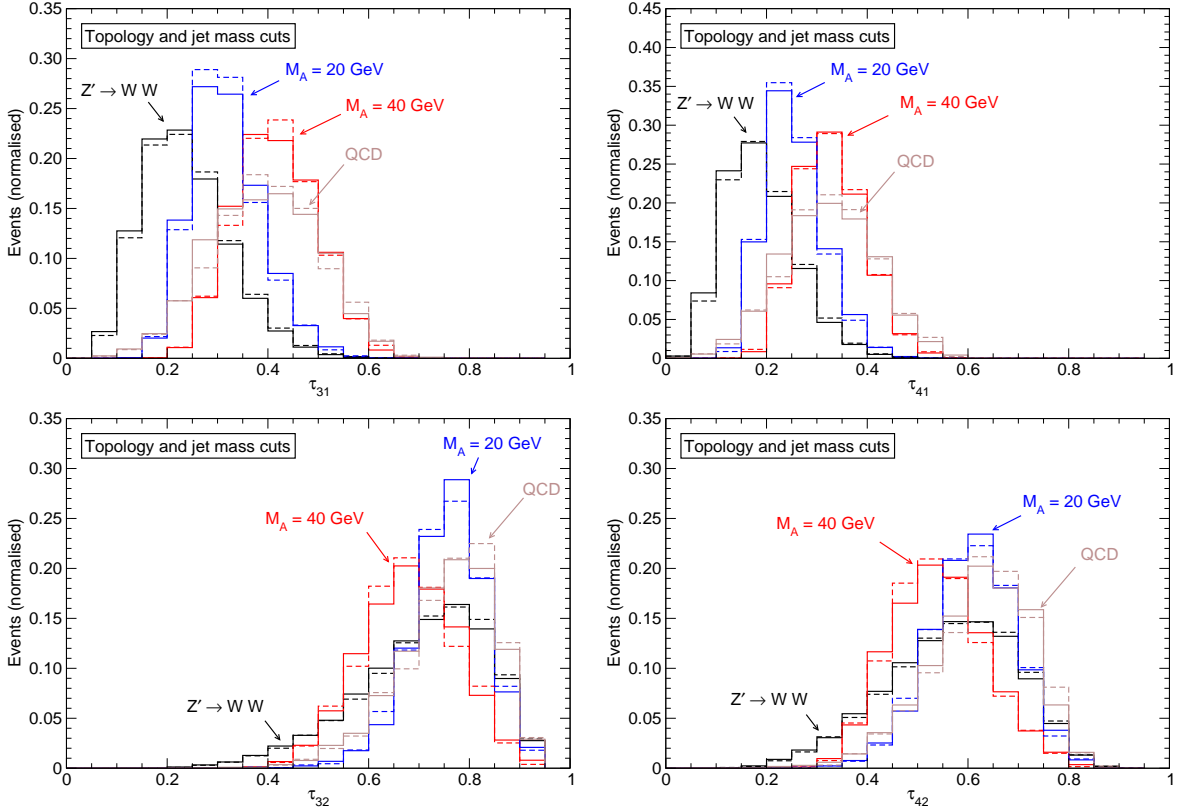

\begin{center}
\begin{tabular}{cc}
\includegraphics[height=5.25cm,clip=]{Figs/tau31-Zp.eps} &
\includegraphics[height=5.25cm,clip=]{Figs/tau41-Zp.eps} \\
\includegraphics[height=5.25cm,clip=]{Figs/tau32-Zp.eps} &
\includegraphics[height=5.25cm,clip=]{Figs/tau42-Zp.eps}
\end{tabular}
\caption{Higher-order $N$-subjettiness ratios $\tau_{nm} = \tau_n / \tau_m$ for $R \to SS$ and $Z' \to W W$ signals and the QCD dijet background. Solid (dashed) lines correspond to the leading (subleading) jet.}
\label{fig:tauN1}
\end{center}
\end{figure}

As we have stressed, the hadronic signals of stealth bosons are hard to spot over the QCD background by using the standard discriminators specifically built for tagging $W$ and $Z$ vector bosons. Still, there are some differences in the jet substructure that can be exploited. As an example, we show in Fig.~\ref{fig:tauN1} several ratios $\tau_{nm} = \tau_n / \tau_m$ for  $Z' \to WW$ and stealth boson signals and the QCD background. We observe that the distributions of $\tau_{32}$ and $\tau_{42}$ are different for $M_A = 40$ GeV, which is the most difficult case, and the background. On the other hand, $\tau_{31}$ and $\tau_{41}$ are different for $M_A = 20$ GeV and the background. With the use of a complete set of $N$-subjettiness observables~\cite{Datta:2017rhs} and a neural network multivariate analysis one can efficiently discriminate these and other multi-pronged signals against the QCD background~\cite{Aguilar-Saavedra:2017rzt}.

We have not addressed the visibility of these signals in the semileptonic diboson searches. Still, a few comments are in order. For $R \to SS$ signals the leptonic decays are absent, and for $R \to VVS$ we expect a similar efficiency suppression for the semileptonic searches as it was found for $VVX$ triboson signals in Ref.~\cite{Aguilar-Saavedra:2016xuc}. The only potentially visible leptonic signatures may arise for $R \to VS$ signals, which deserve a detailed study. For these final states the efficiency is generally expected to be smaller than for true diboson resonances, and with the particularity that the signal may pollute the control regions where the SM backgrounds from $t \bar t$, $W/Z$ plus jets, etc. are normalised, causing an unpredictable effect in the signal regions. This is quite a delicate study and falls out of the scope of this paper. Other signatures from the production of the light states $S$ and $A$ depend on their coupling to SM fermions --- note that they do not couple to SM gauge bosons --- and it is likely that with adequate model building the potential constraints can be evaded.

Our results provide guidance for new hadronic diboson searches. First, a better investigation of the nature of the recurrent excesses found is required. Second, and more importantly, a wider scope of new physics searches beyond the SM is compulsory, as the existing searches may nearly miss these more complex signatures. In particular, (i) generic anti-QCD jet taggers, which are sensitive not only to SM boosted particles but also non-SM ones, should be used;
(ii) relaxing the mass window for at least one of the bosons would allow to investigate signals with non-SM bosons, in particular cascade decays such as in Eq.~(\ref{ec:stealth2}); (iii) in current searches, additional signal regions or analyses with looser requirements on $\tau_{21}$ and $D_2$ should be also considered; (iv) triboson resonance searches should also be performed. In diboson resonance searches, a sufficiently large sample would allow to investigate the profile and characteristics of the deviations --- and this implies that a further tightening of the boson tagging should be avoided.

Finally, let us stress that, beyond diboson resonance searches and the interpretation of anomalies, the stealth boson signals presented here provide a simple new physics case that highlights the limitations of current LHC searches. In order to broaden the sensitivity to new physics signatures, new tools and strategies are needed. First, generic taggers~\cite{Aguilar-Saavedra:2017rzt} that are sensitive to these non-standard boosted signals should be used, at least as an alternative to the dedicated ones. And, in parallel, new grooming algorithms (or variations of existing ones) that correctly recover the mass of multi-pronged objects should also be used.

\section*{Acknowledgements}
I thank J.C. Collins and S. Lombardo for previous collaboration. This work has been supported by MINECO Projects  FPA 2016-78220-C3-1-P and FPA 2013-47836-C3-2-P (including ERDF), and by Junta de Andaluc\'{\i}a Project FQM-101.

\appendix

\section{Test of the background shaping}
\label{sec:a0}

\begin{figure}[htb]
\begin{center}
\includegraphics[height=5.2cm,clip=]{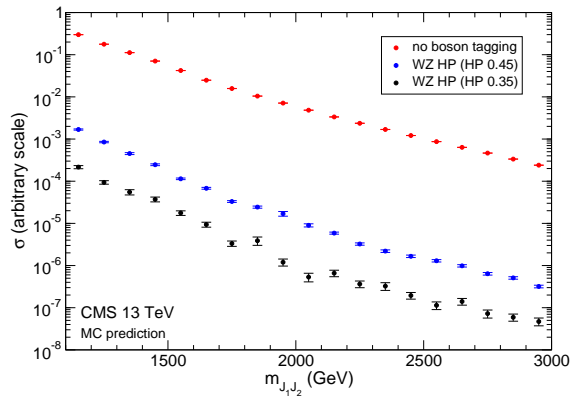}
\caption{Monte Carlo predictions for the QCD dijet background, with the event selection for the CMS Run 2 hadronic diboson searches. The red line corresponds to the topological selection, and the blue and black lines to the $WZ$ HP selection defining HP jets as those with $\tau_{21} \leq 0.45$ or $\tau_{21} \leq 0.35$, respectively.
The error bars in the points represent the Monte Carlo uncertainty. }
\label{fig:QCD}
\end{center}
\end{figure}

In this appendix we investigate a possible shaping of the QCD background due to the CMS event selection criteria, using a high-statistics sample.
We plot in Fig.~\ref{fig:QCD} the dijet invariant mass distribution of the QCD background, after topological selection and also after $WZ$ tagging, either defining HP jets as those with $\tau_{21} \leq 0.45$ or $\tau_{21} \leq 0.35$. 
From the $4.2 \times 10^7$ QCD events simulated, 2587283 events survive the topological selection. The number of events in the HP $WZ$ selection is 7955 (1045) when $\tau_{21} \leq 0.45$ ($\tau_{21} \leq 0.35$) is used to define HP jets. By eye it can be observed that, within the available Monte Carlo statistics, no background shaping is produced. In particular, some tiny bumps in the distribution with the looser selection (blue line), e.g. in the $1.9-2.0$ TeV bin, are not present in the distribution with the harsher selection (black line).

A further check can be done by taking the dijet mass distributions predicted by simulation as pseudo-data, to perform a likelihood test for the presence of bumps over a smooth function parameterised as~\cite{Khachatryan:2014hpa}
\begin{equation}
\frac{dN}{dm_{JJ}} = \frac{P_0 (1 - m_{JJ}/\sqrt s)^{P_1}}{(m_{JJ}/\sqrt s)^{P_2}} \,.
\label{ec:f3P}
\end{equation}
This allows to test whether this functional form adequately parameterises the background, in particular.
The precise details of the procedure applied (which is the standard one to obtain upper limits on a possible narrow resonance signal) are described in Ref.~\cite{Aguilar-Saavedra:2017iso}. We normalise these pseudo-data samples to have the same number of events in the $1.9 - 2.0$ TeV bin as the best-fit background prediction in Ref.~\cite{Sirunyan:2017acf} with 35.9 fb$^{-1}$  (7.9 events). 

\begin{figure}[t]
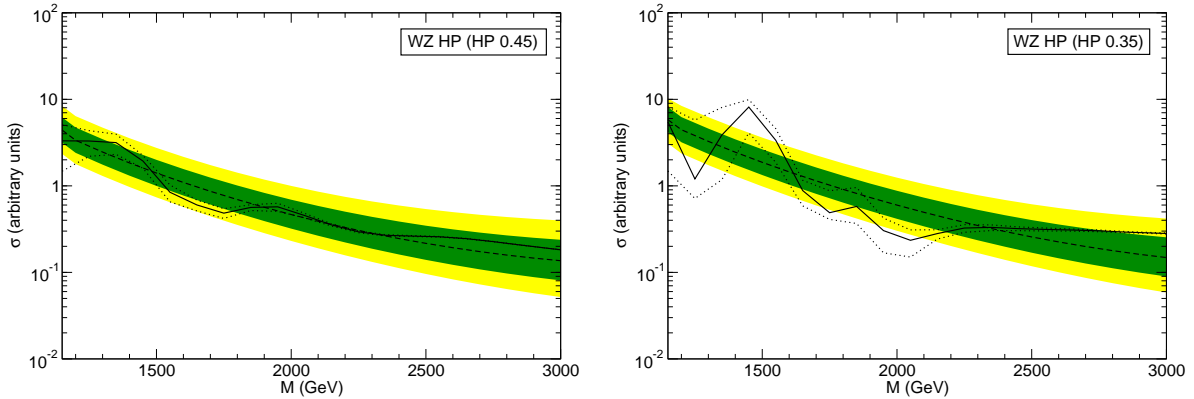

\begin{center}
\begin{tabular}{cc}
\includegraphics[height=5.25cm,clip=]{Figs/QCDtest-45.eps} &
\includegraphics[height=5.25cm,clip=]{Figs/QCDtest-35.eps} 
\end{tabular}
\caption{Upper limits on the cross section of a possible narrow resonance signal, as a function of its mass $M$, obtained using the background distributions in Fig.~\ref{fig:QCD} after selection as pseudo-data.}
\label{fig:QCD-test}
\end{center}
\end{figure}

The expected and observed limits on a possible signal, taking the distributions after selection in Fig.~\ref{fig:QCD} (blue and black lines) as pseudo-data, are presented in Fig.~\ref{fig:QCD-test}. As usual, the dashed lines and green, yellow bands correspond to the expected limits and the $1\sigma$ and $2\sigma$ variation, respectively. Together with the observed limits (solid black lines) we include a simple estimation of their uncertainty arising from the Monte Carlo finite statistics (dotted lines).\footnote{The uncertainty is obtained by varying the data in a single bin within $1\sigma$ of the estimated Monte Carlo uncertainty, keeping the data in the rest of the bins fixed at their central values, and performing the maximum likelihood fit. This uncertainty does not capture the situation where adjacent bins fluctuate in the same direction, in which case the deviation in the observed limits is larger. Thus, the uncertainty is actually underestimated, though for our purposes this approximate assessment suffices.}
For the looser selection (left panel) we see that the dijet mass distribution after simulation is well described by the functional form (\ref{ec:f3P}). For the harsher selection (right panel) we observe some distortions caused by the finite Monte Carlo statistics. For example, in the $1.4 - 1.5$ TeV bin the MC prediction is of $244 \pm 32$ events, with a Monte Carlo uncertainty that is twice larger than the statistical uncertainty. Taking into account the Monte Carlo uncertainty, estimated by the dotted lines, we can see that the variations in the observed limit with respect to the expected one are precisely due to the Monte Carlo statistics, and significant deviations with respect to the assumed functional form (\ref{ec:f3P}) are not observed either.

Finally, let us note that our simulation of the background only includes QCD-mediated dijet production, and not electroweak production of $V$+jets. The inclusion of the latter will likely not change our conclusions, because the $Vj$ invariant mass distribution is also smoothly decreasing at TeV-scale masses, and amounts to a small fraction of the overwhelming QCD background. A shaping of this (smaller) background caused by the jet grooming and event selection, on a final state with a vector boson plus one or more jets, seems unlikely but cannot be discarded with the simulations performed here on a final state with two jets.

\section{Grooming of multi-pronged jets}
\label{sec:a}

Grooming algorithms are designed to remove from hadronic jets the contamination that arises from pile-up, initial state interactions and multiple interactions, trying to recover the mass of the particle originating the jet from the measured jet mass. Although the different algorithms are very useful for $W$ and $Z$ bosons, they are not adequate in general for multi-pronged boosted particles such as stealth bosons. As examples of grooming algorithms, we consider here
\begin{enumerate}
\item The soft-drop algorithm~\cite{Larkoski:2014wba}, which starts from all original constituents of the jet reclustered with the Cambridge-Aachen (CA) algorithm~\cite{Catani:1993hr} and iteratively breaks the jet into two subjets. If the subjets pass the soft-drop condition~\cite{Larkoski:2014wba} with $z_\text{cut} = 0.1$, $\beta = 0$, then the jet is considered as the groomed jet, otherwise the procedure is applied again on the hardest of the two subjets. 
\item Jet trimming~\cite{Krohn:2009th} which reclusters the large-$R$ jet constituents using the anti-$k_T$ algorithm with $R=0.2$ and dropping any of the sub-jets with $p_T$ less than $f_\text{cut} = 0.05$ of the original jet $p_T$.
\item Jet pruning~\cite{Ellis:2009me} which starts from all original constituents of the jet and discards soft recombinations after each step of the CA algorithm. Given two subjets, if their recombination is considered as soft, taking the parameters $z_\text{cut} = 0.1$, $R_\text{cut} = 0.5$, the softer subjet is discarded. 
\end{enumerate}

\begin{figure}[t]
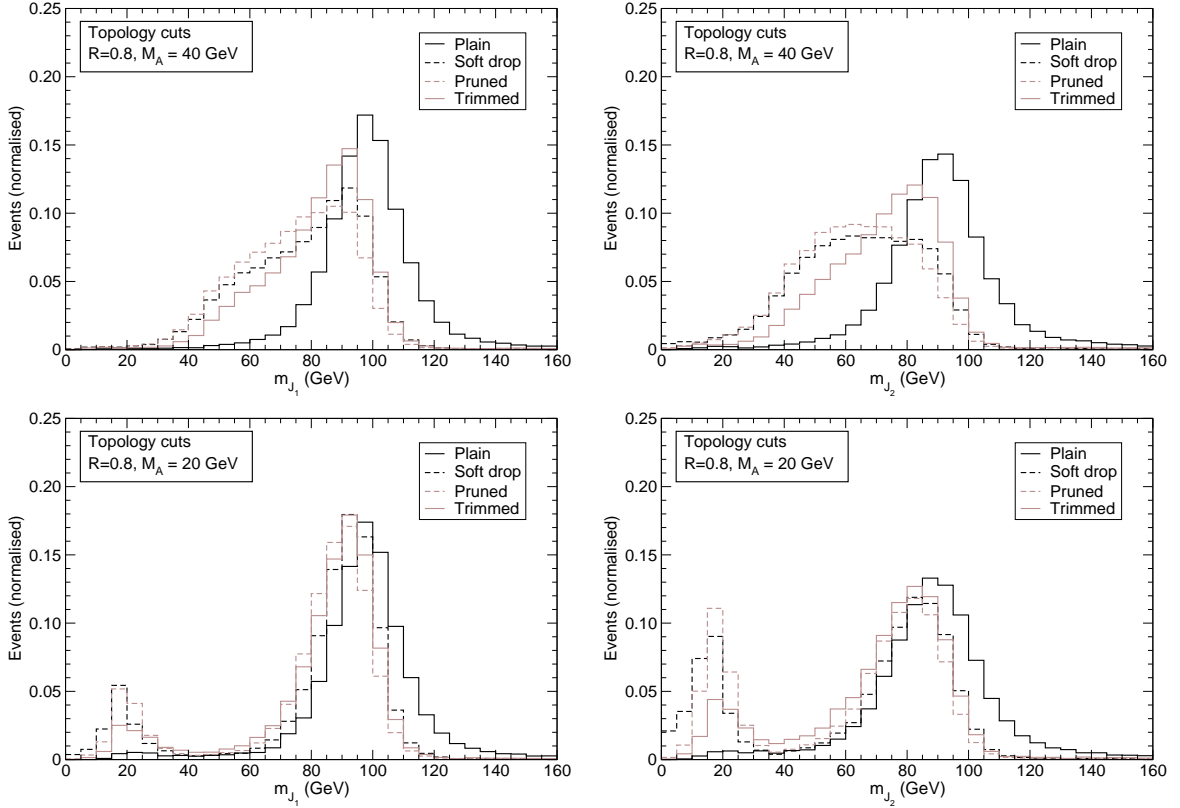

\begin{center}
\begin{tabular}{cc}
\includegraphics[height=5.25cm,clip=]{Figs/mJ1-A40-AK8-comp.eps} &
\includegraphics[height=5.25cm,clip=]{Figs/mJ2-A40-AK8-comp.eps} \\
\includegraphics[height=5.25cm,clip=]{Figs/mJ1-A20-AK8-comp.eps}  &
\includegraphics[height=5.25cm,clip=]{Figs/mJ2-A20-AK8-comp.eps} 
\end{tabular}
\caption{Jet mass distributions for $R \to SS$ and different grooming algorithms, for jet radius $R=0.8$.}
\label{fig:groom08}
\end{center}
\end{figure}

\begin{figure}[t]
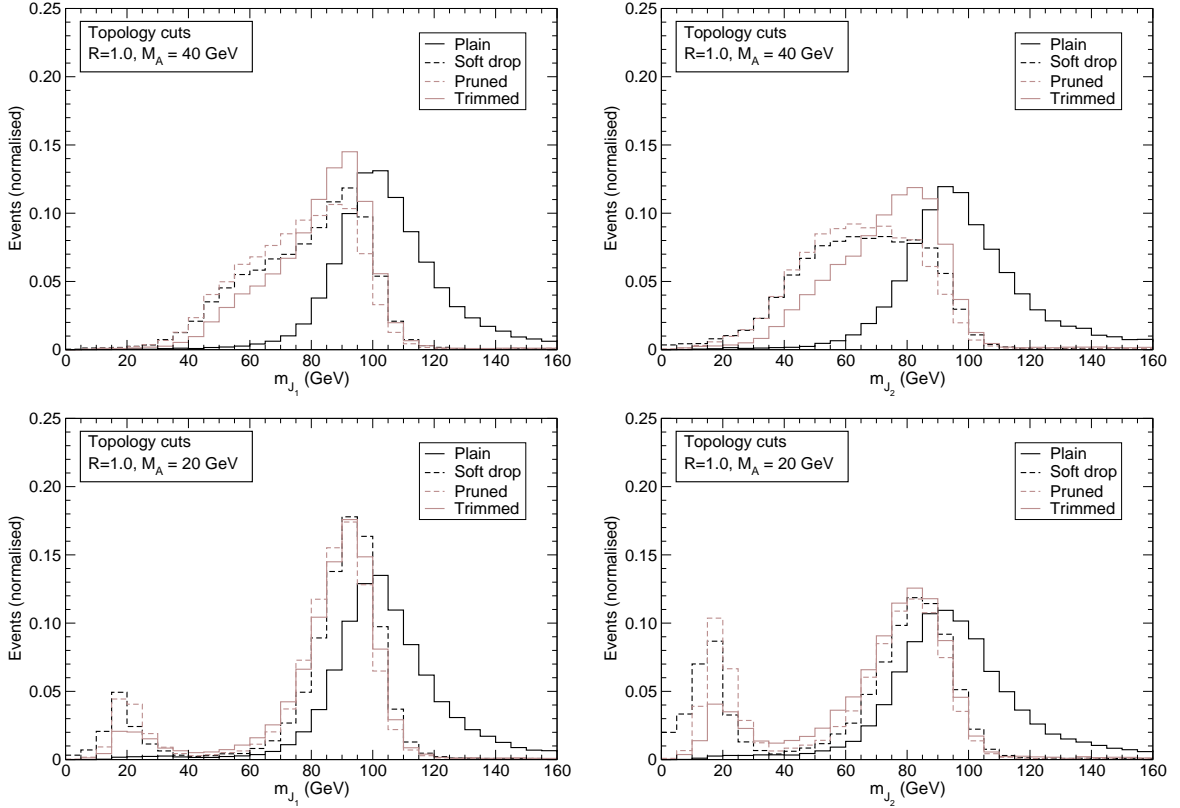

\begin{center}
\begin{tabular}{cc}
\includegraphics[height=5.25cm,clip=]{Figs/mJ1-A40-AK10-comp.eps} &
\includegraphics[height=5.25cm,clip=]{Figs/mJ2-A40-AK10-comp.eps} \\
\includegraphics[height=5.25cm,clip=]{Figs/mJ1-A20-AK10-comp.eps}  &
\includegraphics[height=5.25cm,clip=]{Figs/mJ2-A20-AK10-comp.eps} \\ 
\end{tabular}
\caption{Jet mass distributions for $R \to SS$ and different grooming algorithms, for jet radius $R=1.0$.} 
\label{fig:groom10}
\end{center}
\end{figure}

We present in Fig.~\ref{fig:groom08} the jet masses for the $R \to SS$ signals in section~\ref{sec:2}, for AK8 jets and with the topology cuts of the CMS analyses. We also show the plain jet mass for comparison. As we can see, for $M_A = 40$ GeV (top panels) the jet mass distribution is considerably spoiled by the grooming: the mass peak in the plain mass distribution is not sharpened but, on the contrary, it is transformed into a wide bump. For $M_A = 20$ GeV the substructure of the jets is more two-prong-like (as it was also seen in section~\ref{sec:2} by considering the $\tau_{21}$ variable) and the groomed mass is closer to the plain mass. It is interesting to note that a small mass peak appears at 20 GeV, when the grooming algorithms completely remove one of the $S \to AA$ decay products --- see also appendix~\ref{sec:b} for the analogous case of heavier stealth bosons. For wider jets with $R=1.0$ the results are alike, see Fig.~\ref{fig:groom10}. The wider jet radius makes the jets catch a larger amount of contamination, and this is reflected in a larger high-mass tail of the plain mass distribution. This contamination is removed by the grooming but, at the same time, the peaks are distorted, especially for $M_A = 40$ GeV, for which the structure of the jets departs more from a two-pronged decay. These results highlight the need to tune a grooming algorithm that correctly removes contamination from multi-pronged non-SM boosted objects, which is beyond the scope of the present work.

\section{Heavier stealth bosons}
\label{sec:b}

Throughout this work we have studied stealth bosons with a mass $M_S = 100$ GeV, which we have selected close to the $W$ and $Z$ masses, in order to investigate how the signature of these elusive particles would appear in diboson resonance searches. From our results, it is easy to realise that heavier boosted particles also giving rise to a four-pronged jet will share the same behaviour in which regards the jet substructure. (See also Ref.~\cite{Aguilar-Saavedra:2017rzt}.) Here we only address, for completeness, the decay chain in Eq.~(\ref{ec:stealth2}) where the jet grooming may yield a jet mass close to the $Z$ mass. Heavier stealth bosons could be produced in pairs, or associated to one or two weak bosons, in much the same way as the lighter ones studied in sections~\ref{sec:2}, \ref{sec:3} and \ref{sec:4}, respectively. For brevity we will not repeat all the analyses in those sections for this case, but will restrict ourselves to the CMS event selection and show that the decay chain of Eq.~(\ref{ec:stealth2}) can produce jets with a groomed mass $m_J \sim M_V$ and whose substructure is not seen as two-pronged by the $\tau_{21}$ discriminator.

We consider $Z' \to S Z$, in which we select the decay $Z \to \nu \bar \nu$ in order to clearly identify the fat jet from the stealth boson decay, and $S \to ZA$. As before, we set $M_{Z'} = 2$ TeV and for the heavier stealth boson we select a mass $M_S = 200$ GeV. The analysis is done applying the event selection criteria of CMS diboson searches detailed in section~\ref{sec:2}.

\begin{figure}[t]
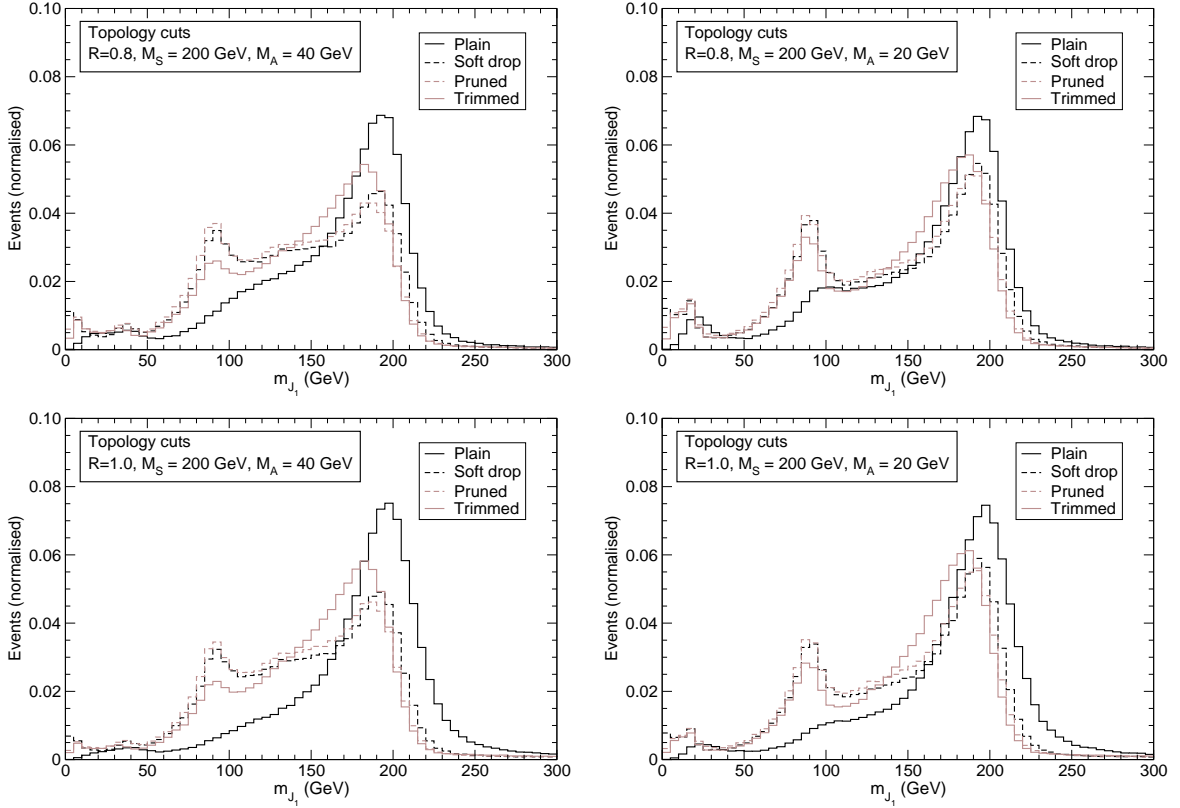

\begin{center}
\begin{tabular}{cc}
\includegraphics[height=5.25cm,clip=]{Figs/mJ-S200-A40-AK8-comp.eps} &
\includegraphics[height=5.25cm,clip=]{Figs/mJ-S200-A20-AK8-comp.eps} \\
\includegraphics[height=5.25cm,clip=]{Figs/mJ-S200-A40-AK10-comp.eps}  &
\includegraphics[height=5.25cm,clip=]{Figs/mJ-S200-A20-AK10-comp.eps} 
\end{tabular}
\caption{Jet mass distributions for $S \to ZA$ and different grooming algorithms, for jet radii $R=0.8$ (top) and $R=1.0$ (bottom).}
\label{fig:groomH}
\end{center}
\end{figure}

The leading jet mass, without grooming and after application of various grooming algorithms, is presented in Fig.~\ref{fig:groomH} for $M_A = 40$ GeV (top, left) and $M_A = 20$ GeV (top, right). We first notice that the plain jet mass has a low-mass tail, originated because sometimes the AK8 jets do not contain all the $S$ decay products. With a wider jet radius $R=1.0$ (bottom panels) this effect is softened. Of course, for a larger $M_{Z'}$ the fat jets would be more collimated, but for better comparison with the results in the rest of the paper we have kept a 2 TeV heavy resonance mass. From these plots we again observe that, as seen for $M_S = 100$ GeV in the previous appendix, the grooming procedure significantly modifies the jet mass distribution. For soft drop and pruning, the appearance of a peak at the $Z$ mass, nearly as high as the peak at $M_S$, is remarkable. This happens when the grooming removes all the decay products of $A$ from the jet. Smaller peaks at $M_A$ are also visible. Therefore, we see that a heavier particle of twice the $Z$ boson mass can often yield a jet with groomed mass close to the weak boson masses.

\begin{figure}[t]
\begin{center}
\includegraphics[height=5.25cm,clip=]{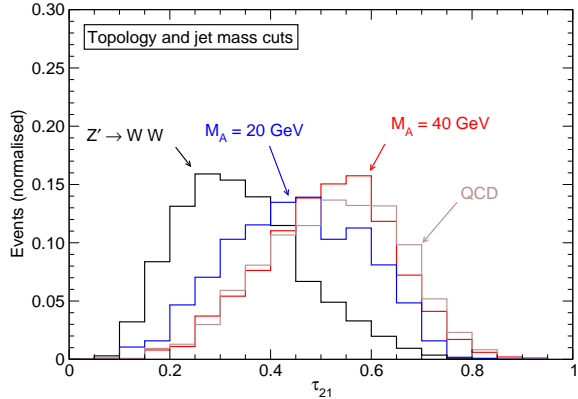}
\caption{$N$-subjettiness ratio $\tau_{21}$ for heavier stealth boson signals $S \to ZA$. For comparison, we show the distributions for the leading jet in $Z' \to W W$ and the QCD dijet background.}
\label{fig:XX-heavy}
\end{center}
\end{figure}

The $\tau_{21}$ variable of events passing the jet mass cut $m_J \in [65,105]$ GeV is presented in Fig.~\ref{fig:XX-heavy}. The distribution for fat jets from $S \to ZA$ with $M_A = 40$ GeV is quite close to that of QCD events, while for $M_A = 20$ GeV it is slightly shifted towards smaller $\tau_{21}$ values. Therefore, we see that these particles behave as stealth bosons, when one considers variables measuring the two-pronged structure. Without the jet mass cut, the distributions are found to be comparable.

In summary, the decay chain in Eq.~(\ref{ec:stealth2}) would lead to the very conspicuous signal of a heavier new particle of a few hundreds of GeV, with hadronic four-pronged decay --- then giving a jet with a substructure quite different from that of $W$, $Z$ bosons --- but with a groomed jet mass that often is close to the weak boson masses. Obviously, such a signal would be penalised in diboson resonance searches, and removing the jet mass cut would lead to an enhancement of signal, and also of the background. Semi-leptonic signals of these particles, produced when the $Z$ boson decays leptonically, are likely to be highly suppressed by the isolation requirement on charged leptons, as these are very close to the hadronic decay products of $A$. This type of signal constitutes another new physics case that calls for novel tools that can correctly identify non-SM boosted jets.

\end{document}